%% file: paper_revised_r2.tex
\documentclass[journal]{IEEEtran}
\usepackage{graphicx, tikz} % Required for inserting images
\usepackage{amsmath, amssymb, amsthm, bbm}
\usepackage{subcaption}
\usepackage{float}
\usepackage{xcolor}
\usepackage{comment}
\usepackage{caption}
\title{Rate-Splitting Multiple Access for Integrated Sensing and Communications: A First Experimental Study}
\author{Xinze Lyu, Sundar Aditya, \IEEEmembership{Member, IEEE} and Bruno Clerckx, \IEEEmembership{Fellow, IEEE}
\thanks{This work was partially funded by UKRI Impact Acceleration Account (IAA) grant EP/X52556X/1 and UKRI grants EP/X040569/1, EP/Y037197/1, EP/X04047X/1, EP/Y037243/1.}
\thanks{X.~Lyu and B.~Clerckx are with the Dept.~of Electrical and Electronic Engg., Imperial College London, London SW7 2AZ, U.K. (e-mail: \{x.lyu21, b.clerckx\}@imperial.ac.uk).}
\thanks{S. Aditya is with Meobyr, London EC1V 2NX, U.K. (e-mail: s.aditya@imperial.ac.uk).}
}

\include{notation_new}

\begin{document}
\maketitle

\bstctlcite{IEEEexample:BSTcontrol}
\begin{abstract}
 A canonical use case of Integrated Sensing and Communications (ISAC) in multiple-input multiple-output (MIMO) systems involves a multi-antenna transmitter communicating with $K$ users and sensing targets in its vicinity. For this setup, precoder and multiple access designs are of utmost importance, as the limited transmit power budget must be efficiently directed towards the desired directions (users and targets) to maximize both communications and sensing performance. This problem has been widely investigated analytically under various design choices, in particular (a) whether or not a dedicated sensing signal is needed, and (b) for different MIMO multiple access techniques, such as Space Division Multiple Access (SDMA) and Rate-Splitting Multiple Access (RSMA). However, a conclusive answer on which design choice achieves the best ISAC performance, backed by experimental results, remains elusive. We address this vacuum by experimentally evaluating and comparing RSMA and SDMA for communicating with two users $(K = 2)$ and sensing (ranging) one target. Over three scenarios that are representative of \emph{vehicular} ISAC, covering different levels of inter-user interference and separation/integration between sensing and communications, we show that RSMA without a dedicated sensing signal achieves better ISAC performance -- i.e., higher sum throughput (up to $50\%$ peak throughput gain) for similar radar SNR (between $20$ to $24{\rm dB}$) -- than SDMA with a dedicated sensing signal. This first-ever experimental
study of RSMA ISAC demonstrates the feasibility and the superiority of RSMA for future multi-functional wireless systems.
\end{abstract}

\section{Introduction}
\label{sec:intro}
Integrated Sensing and Communications (ISAC) is widely anticipated to be a prominent feature of future wireless networks, as evidenced by standardization activities on ISAC for 6G \cite{3gpp_isac}. One strand of ISAC that has been extensively investigated by the research community is multiple-input multiple-output (MIMO) ISAC, as the MIMO paradigm is mature in both communications (comms) and radar \cite{Liu_jsac_survey}. In particular, a problem that has been widely investigated involves a multi-antenna transmitter (TX) simultaneously serving $K$ communications users (UEs) and sensing targets in its vicinity. For this problem, the communications and sensing performance is closely linked to the amount of power radiated towards the UEs and targets, respectively. As a result, the choice of precoders (beamforming) and multiple access techniques used at the TX has a direct bearing on ISAC performance. Indeed, in the context of ISAC, multiple access techniques make use of the
resource dimensions (e.g., time, frequency, power, antenna, code,
message, etc) to serve multiple physical users (communications) and virtual users (target to sense), ideally in the most efficient way \cite{PIEEE_bruno_2024_ma4intelligent6g}.

% Literature review
The ISAC precoder design problem has been extensively studied to jointly optimize communication and sensing performance metrics in various well-motivated scenarios -- conventional MIMO with fully digital beamforming \cite{liuXiang_TSP_MIMOprecoder, Liu_TSP_MinCRB_QoS, Lu_TSP_dedicated_required, cui_wcnc_MSE_DoF,Liu_erzi_ISAC_SI_JSAC_2023}, massive MIMO \cite{Fan_mMIMO_isac_iot_2024,liao_mMIMO_isac_twc_2024,topal_mMIMO_isac_wcml_2024,demirhan_ML_isac_mMIMO_spawc_2024}, vehicular systems \cite{Liu_TWC_radar_predict_vehiculat, Yuan_TWC_V2I_vehicular}, millimeter-wave with hybrid beamforming \cite{qi_TCOML_ISAC_HBF_mmWave, wang_TCOM_2022_ISAC_HybridBF,gupta_mmwave_ISAC_OJ_2024,wang_mmwave_isac_iot_2024}, using low resolution analog-to-digital converters \cite{Lin_mmwave_isac_onebit_2024}, etc. There are two implicit assumptions across these works. Firstly, they can all be viewed as adapting Space Division Multiple Access (SDMA) -- the prevailing multiple access technique used in 5G -- for ISAC, and secondly, they either assume a dedicated sensing signal \cite{Fan_mMIMO_isac_iot_2024,demirhan_ML_isac_mMIMO_spawc_2024,topal_mMIMO_isac_wcml_2024, qi_TCOML_ISAC_HBF_mmWave}, or reuse communications signals\footnote{By this, we mean the data payload comprising channel-coded and modulated symbols.} for sensing \cite{Liu_TSP_MinCRB_QoS, cui_wcnc_MSE_DoF,Liu_erzi_ISAC_SI_JSAC_2023,liao_mMIMO_isac_twc_2024, Liu_TWC_radar_predict_vehiculat, Yuan_TWC_V2I_vehicular, wang_TCOM_2022_ISAC_HybridBF,gupta_mmwave_ISAC_OJ_2024,wang_mmwave_isac_iot_2024,Lin_mmwave_isac_onebit_2024} as pre-fixed design choices without scrutinizing the merits of each choice. As a result, two fundamental questions can be posed w.r.t the ISAC precoder design problem, namely:
\begin{itemize}
    \item[Q1.] Are there alternatives to SDMA that could provide better ISAC performance\footnote{See Definition~\ref{def:isac_perf_region} on page 6 for our definition of ISAC performance.}?
    \item[Q2.] Are separate signals needed for communications and sensing for the best ISAC performance?
\end{itemize}
These are especially pertinent questions, as specifications related to the waveform and multiple access technique are among the first to be decided in every new generation of standards.

Regarding Q1, Rate-Splitting Multiple Access (RSMA), which generalizes SDMA (and NOMA), has been shown both analytically and empirically to outperform SDMA (and NOMA) w.r.t communications \cite{lyu2023prototype, RSMA_MGM_prototype,RSMA_JSAC_Primer}. Briefly, using RSMA to communicate with $K$ UEs involves the design of $K+1$ precoders to transmit $K+1$ communications signals \cite{RSMA_JSAC_Primer}. Likewise, using SDMA to communicate with $K$ UEs and track a single target using a dedicated sensing signal also involves the design of $K+1$ precoders\footnote{Essentially, the target can be viewed as a virtual UE.}. These similarities strongly motivate an RSMA v/s SDMA ISAC comparison.

As for Q2, it is important to clarify what sensing entails. For instance, if the sensing task involves detecting the presence/absence of a target or direction finding, then clearly these can be accomplished using communications signals with an energy detection receiver or subspace methods like MUSIC, respectively. Importantly, the sensing signal processing for these tasks does not depend on the \emph{content} of the communications signal. For target ranging (i.e., distance estimation) on the other hand, the signal content impacts estimation performance as the range estimator is a matched filter. Since the answer to Q2 is non-trivial for ranging, we choose target ranging as the sensing task in this paper, specifically, ranging based on delay-domain detection. We infer the target’s range by observing the peak of the post-processing radar SNR across discrete delay bins. This process implicitly performs both target detection (presence) and ranging (location), and reflects a practical method for estimating delay in indoor ISAC systems. However, it has been shown that the ranging performance of communications signals matches that of dedicated sensing signals -- e.g., frequency modulated continuous wave (FMCW) -- asymptotically at large signal lengths \cite{Adi_ccs_ojcoms_2023}. 

Taken together, Q1 and Q2 motivate the following four-way ISAC comparison -- RSMA/SDMA with/without dedicated sensing signal. This has been investigated in \cite{XuCC_JTSP_RSMA_ISAC, Liu_ICC_RSMA_ISAC_2023, Liu_TWC_RSMA_ISAC_sate, loli2022ratesplitting,YinLF_CommL_RSMA_ISAC_wo} with link-level simulations indicating that RSMA without a dedicated sensing signal achieves the best ISAC performance -- intuitively, the \emph{extra} precoder (i.e., $K+1$ precoders for $K$ UEs) can be used for sensing\footnote{This works only because the communications signal carried by the precoder can also be used for sensing. Thus, Q1 and Q2 in page 1 are closely related questions.}. In particular, \cite{XuCC_JTSP_RSMA_ISAC} was the first to theoretically show that RSMA can be used for the dual purpose of boosting the communication sum rate of multiple users and simultaneously increasing the sensing performance, hence enlarging the communication-sensing performance trade-off. This is due to the common streams in RSMA that can be used to manage inter-user interference and sensing targets. However, a conclusive answer on which design choice achieves the best ISAC performance, backed by experimental results, is missing in the literature -- experimental evaluations of ISAC for single-antenna systems is presented in \cite{ISAC_SISO_proto_1, ISAC_SISO_proto_2, ISAC_SISO_proto_otfs} and for multi-antenna ISAC (SDMA) in \cite{Xu_OJ_2022_ISAC_MIMO_proto_beamform}. In this paper, we address this void, and our contributions are as follows:
\begin{itemize}
    \item We formulate a signal model for RSMA ISAC with a dedicated sensing signal, which includes the other design choices (SDMA ISAC, RSMA ISAC without a dedicated sensing signal) as special cases. We then evaluate ISAC performance using two key metrics: (i) sum throughput (communication metric), and (ii) post-processing radar SNR, which reflects the strength of the radar return signal used for target ranging. These metrics provide a practical and quantifiable basis for comparing different ISAC design choices. A formal definition of the \emph{ISAC performance region} based on these metrics is provided in Section~\ref{sec:ISAC metric} Definition \ref{def:isac_perf_region}.
    
    \item Using software-defined radios (SDRs), we implement the above signal model using OFDM signals largely based on IEEE 802.11ac-VHT physical layer frames \cite{IEEE_80211_ac}. We consider a signal bandwidth of $100{\rm MHz}$, yielding a range resolution (bin size) of $1.5{\rm m}$.
    
    \item In our measurements, we consider three scenarios that are representative of \emph{vehicular ISAC}\footnote{While the target is stationary in our testbed, it is selected to emulate the radar cross-section of a vehicle.}, covering different levels of inter-user interference and separation/integration between sensing and communications.  Over these scenarios, we observe that:
    \begin{itemize}
        \item[a)] The SDMA ISAC performance boundary lies in the interior of the RSMA ISAC performance region. Hence, RSMA ISAC outperforms SDMA ISAC. 

        \item[b)] Moreover, the RSMA ISAC performance boundary is achieved \emph{without a dedicated sensing signal}. In contrast, a dedicated sensing signal is needed to achieve the SDMA performance boundary. 

        \item[c)] The gap between the SDMA and RSMA ISAC performance regions is scenario dependent. In particular, RSMA without a dedicated sensing signal can achieve peak sum throughput gains of up to $50\%$ over SDMA with a dedicated sensing signal, for similar radar SNR (between $20$ and $24{\rm dB}$). 
    \end{itemize}

\end{itemize}

\subsection{Organization}
The rest of this paper is organized as follows. In Section~\ref{sec:sysmodel}, we describe the signal model for RSMA ISAC with a dedicated sensing signal; the other design choices are special cases of this general case (see Section \ref{subsec:splcases}). In Section~\ref{sec:rsma isac prototype}, we present details of our SDR-based RSMA ISAC prototype, specifically the control signaling, OFDM payload structure, MCS levels etc. In Section~\ref{subsec: meas setup}, we motivate our measurement scenarios that are representative of vehicular ISAC. Capturing the peak performance of RSMA ISAC with a dedicated sensing signal involves a four-dimensional parameter search, which is cumbersome to realize in an experimental setting. This motivates us to pursue \emph{simulation-aided parameter search}, described in Section~\ref{subsec:sim_param_search}, whereby we obtain a subset of well-chosen parameters that we use to measure and compare the ISAC performance of RSMA and SDMA, the results of which are presented in Section~\ref{subsec:measured_isac_perf}. Finally, Section~\ref{sec: concl} concludes the paper.

\subsection{Notation}
Column vectors are represented by lowercase bold letters (e.g., $\mathbf{h}$). $\mathbf{1}$ denotes the all-one vector, $|\cdot|$ the magnitude of scalars, $(\cdot)^H$ the Hermitian operator, $\|\cdot\|$ the Euclidean norm, and $\mathcal{CN}(0,\sigma^2)$ the circularly symmetric complex Gaussian distribution with zero mean and variance $\sigma^2$. For a uniform linear array with $N_T$ antenna elements and half-wavelength spacing, $\nba_{\theta} = [1~ e^{j\pi \sin \theta} ~ \cdots ~ e^{j \pi (N_T-1) \sin \theta}]^T$ denotes the steering vector along direction $\theta$.

\section{System Model}
\label{sec:sysmodel}
Consider a TX with $N_T$ antennas serving two single-antenna UEs and sensing (target ranging based on delay-domain detection) one target at its broadside as a monostatic radar in tracking mode. {The target in our experimental setup is physically stationary, it is selected to emulate the radar cross-section of a typical vehicular object. This setup allows for controlled reproducibility of channel conditions.} We consider OFDM signals over $N_c$ subcarriers for both communications and sensing. 

\subsection{Transmit signal}
\label{subsec:tx_signal}
\paragraph{RSMA communications}
For the two-UE case, RSMA communications involves the transmission of three linearly precoded signals \cite{RSMA_JSAC_Primer}. Specifically, let $W_i$ denote the message meant for UE~$i~(=1,2)$. At the TX, each $W_i$ is split into common and private portions -- denoted by $W_{c,i}$ and $W_{p,i}$, respectively. The common portions, $W_{c,1}$ and $W_{c,2}$, are combined into a common message, which is then encoded and modulated to form a \emph{common stream}, $s_c[k]~(k = 0,\cdots,N_c-1)$ over the subcarriers. Similarly, $W_{p,1}$ and $W_{p,2}$ are individually encoded and modulated to form \emph{private streams}, $s_1[k]$ and $s_2[k]$, respectively. We assume zero-mean, unit-energy symbols for each stream. The three streams are then linearly precoded before transmission.

\begin{nrem}[SDMA as a special case of RSMA]
\label{rem:sdma_rsma}
    In the absence of message splitting (i.e., no $s_c$), RSMA reduces to SDMA.
\end{nrem}

\paragraph{Sensing (Ranging)}
Ranging as a stand-alone functionality can be realized by transmitting a deterministic \emph{sensing signal}, $s_r[k]$, aimed at the target using a precoder (beamformer) $\nbp_r$.

\paragraph{RSMA ISAC}
From a) and b) above, the RSMA ISAC transmit signal, $\nbx[k]$, can be modeled as follows:
\begin{align}
\label{eq:x}
    \nbx[k] &= \nbp_c[k] s_c[k] + \nbp_1[k] s_1[k] + \nbp_2[k] s_2[k] + \nbp_r[k] s_r[k],
\end{align}
where for subcarrier $k$, $\nbp_c[k]$ is referred to as the \emph{common stream precoder}, $\nbp_i[k]$ the \emph{private stream precoder} for UE~$i$, and $\nbp_r[k]$ the \emph{sensing precoder}. We assume a sum transmit power constraint, i.e., 
\begin{align}
\sum_{k=0}^{N_c - 1} (\|\nbp_c[k]\|^2 + \|\nbp_1[k]\|^2+\|\nbp_2[k]\|^2+\|\nbp_r[k]\|^2) = P_T    
\end{align}
.
\begin{nrem}[Ranging using communications signals]
\label{rem:ccs}
  To use the communications signals -- $s_c[\cdot]$, $s_1[\cdot]$ and $s_2[\cdot]$ -- for ranging \cite{Adi_ccs_ojcoms_2023}, the respective precoders -- $\nbp_c[\cdot]$, $\nbp_1[\cdot]$ and $\nbp_2[\cdot]$ -- should strike a balance between radiating power towards the desired UE(s) and the target. This trade-off motivates our precoder design choices in the next subsection.
\end{nrem}

\subsection{Precoder Design}
\label{subsec:precoder_design}
Let $\nbh_i[k] \in \nbbC^{N_T}$ denote the slowly varying, flat fading (communications) channel between the TX and UE $i~(=1,2)$. Through downlink pilot signals, we assume UE~$i$ obtains an estimate of $\nbh_i[k]$, denoted by $\hat{\nbh}_i[k]$, which is fed back to the TX. Let $\nbu_i[k] := \hat{\nbh}_i[k]/\|\hat{\nbh}_i[k]\|$ denote the unit vector along $\hat{\nbh}_i[k]$. Similarly, let $\nbu_0 := \nba_0/\sqrt{N_T}$ denote the normalized steering vector along the TX's broadside, which is assumed to be the target direction.

%To reduce channel state information (CSI) overhead, UE $i$ feeds back to the TX the \emph{wideband CSI}, obtaining by averaging $\nbh_i[k]$ over the subcarriers, i.e.,
%\begin{equation}
%    \label{eq:CSIT averaging}
%    \hat{\nbh}_i := \frac{1}{N_c}\sum^{N_c-1}_{k=0}\hat{\nbh}_i[k].
%\end{equation}

\paragraph{Design of $\nbp_c[\cdot]$}
In RSMA communications, each UE independently decodes $s_c[\cdot]$ first by treating the interference from $s_1[\cdot]$ and $s_2[\cdot]$ as noise. Hence, $\nbp_c[\cdot]$ must have sufficient gain at \emph{both} UEs (for communications), as well as the target direction (for sensing, see Remark~\ref{rem:ccs}). This motivates a weighted maximum ratio transmission (MRT) precoder choice for $\nbp_c[\cdot]$, as follows:
\begin{align}
\label{eq:pc}
    \nbp_c[k] &= \sqrt{P_T t_{\rm comms} (1-t_p)} \notag \\
    ~&\times \frac{(\sqrt{\alpha_c} \nbu_c[k] + \sqrt{1 - \alpha_c} \nbu_0)}{\sqrt{\displaystyle\sum\limits_{k'=0}^{N_c-1}\|\sqrt{\alpha_c} \nbu_c[k'] + \sqrt{1 - \alpha_c} \nbu_0\|^2}},
    \end{align}
    where
    \begin{align}
        \label{eq:uc}
\nbu_c[k] &= (\nbu_1[k] + \nbu_2[k])/\|\nbu_1[k] + \nbu_2[k]\|.
    \end{align}

In (\ref{eq:uc}), through equi-weighted MRT along $\hat{\nbh}_1[k]$ and $\hat{\nbh}_2[k]$, the vector $\nbu_c[k]$ ensures some gain at both UEs, and the form of (\ref{eq:pc}) further ensures some gain along the target direction ($\nbu_0$) as well. The parameter $\alpha_c \in [0,1]$ controls the priority between communication vector ($\mathbf{u}_c$) and target direction steering vector ($\mathbf{u}_0$) by forming a weighted combination of those two vectors. Finally, $t_{\rm comms} \in [0,1]$ captures the fraction of the transmit power allocated for communications, and $t_p \in [0,1]$ represents the fraction of $t_{\rm comms}$ that is allocated to the private stream precoders.

\paragraph{Design of $\nbp_i[\cdot]~(i = 1,2)$}
After decoding $s_c[\cdot]$, UE~$i$ subtracts its contribution (i.e., through successive interference cancellation) and then decodes $s_i[\cdot]$ by treating the interference from the other private stream as noise. In our experiments, we consider two standard precoding strategies for SDMA baseline and RSMA private streams: MRT and zero-forcing (ZF).
\begin{itemize}
    \item In the MRT precoder design, the communication vector $\nbu_i[k] = \hat{\nbh}_i[k]/\|\hat{\nbh}_i[k]\|$, which allows the transmit beams to align with the individual users.
    \item In the ZF precoder design, the $\nbu_i[k]=\nbU_k(:,i)$, where $\nbU_k=\nbH_k(\nbH^H_k\nbH_k)^{-1}$ and $\nbH_k=[\hat{h}_1[k],\hat{h}_2[k]]$. By using ZF precoder design, users are nulling each other's inter-user interference.
\end{itemize}
Furthermore, $\nbp_i[\cdot]$ must have sufficient gain at UE~$i$ as well as the target direction. This motivates the following design for $\nbp_i[\cdot]$ similar to (\ref{eq:pc}):

\begin{align}
\label{eq:pi}
    \nbp_i[k] &= \sqrt{\frac{P_T t_{\rm comms} t_p}{2}} \times \frac{(\sqrt{\alpha_p} \nbu_i[k] + \sqrt{1 - \alpha_p}\nbu_0)}{\sqrt{\displaystyle\sum\limits_{k'=0}^{N_c-1} \|\sqrt{\alpha_p} \nbu_i[k'] + \sqrt{1 - \alpha_p} \nbu_0\|^2}} , 
\end{align}
where like $\alpha_c$ in (\ref{eq:pc}), $\alpha_p \in [0,1]$ also allows the beamformer to be a weighted combination of the communication vector ($\mathbf{u}_i$) and target direction steering vector ($\mathbf{u}_0$), which also captures the priority/weight assigned to communications. In general, $\alpha_c \neq \alpha_p$. We further assume that the total private stream precoder power ($= t_{\rm comms} t_p$) is equally divided between $\nbp_1[\cdot]$ and $\nbp_2[\cdot]$. This is a reasonably good choice when the UEs have similar channel strengths (i.e., $\|\hat{\nbh}_1\| \approx \|\hat{\nbh}_2\|$), which is the scenario we focus on in our measurements.

\paragraph{Design of $\nbp_r[\cdot]$}
Since $s_r[\cdot]$ is not an information-bearing signal, $\nbp_r[\cdot]$ needs to have high gain only along the target direction. From (\ref{eq:pc}) and (\ref{eq:pi}), $1-t_{\rm comms}$ is the fraction of the transmit power allocated solely for sensing. Hence, 
\begin{align}
\label{eq:pr}
    \nbp_r[k] &= \sqrt{\frac{P_T (1 - t_{\rm comms})}{N_c}} \times \nbu_0.
\end{align}

\begin{nrem}
    \label{rem:4params}
    Four parameters are used to capture the full extent of the communications-sensing trade-off in (\ref{eq:pc})-(\ref{eq:pr}). In summary, they are:
    \begin{itemize}
        \item $t_{\rm comms} \in [0,1]$: Fraction of transmit power allocated for communications. $t_{\rm comms} \rightarrow 0/1$ implies most of the power allocated for sensing/communications.

        \item $t_p \in [0,1]$: Fraction of communications power (i.e., $t_{\rm comms}$) allocated to the private streams allocated for communications. $t_p \rightarrow 0/1$ implies most of $t_{\rm comms}$ is allocated to the common/private stream(s).

        \item $\alpha_c \in [0,1]$: As $\nbp_c[\cdot]$ can be used for both communications and sensing whenever it is allocated non-zero power, $\alpha_c$ is the priority given to communications. In particular, $\alpha_c \rightarrow 0/1$ implies very high priority given to sensing/communications.

        \item $\alpha_p \in [0,1]$: Like $\alpha_c$, $\alpha_p$ is the priority given to communications for the private stream precoders, $\nbp_1[\cdot]$ and $\nbp_2[\cdot]$. Again, $\alpha_p \rightarrow 0/1$ implies very high priority given to sensing/communications.
    \end{itemize}
\end{nrem}

\begin{nrem}[Heuristic precoders]
    \label{rem:heuristic_precoders}
     The precoders in (\ref{eq:pc})-(\ref{eq:pr}) are sub-optimal\footnote{The heuristic precoders are adaptations driven by specific design inspirations. First, the weighted precoder direction design (the linear combination of $u_c/u_i$ and $u_0$) is inspired by theoretical works \cite{XuCC_JTSP_RSMA_ISAC, loli2022ratesplitting}. Second, the weighted precoder power design (parameters $t_{comms}$ and $t_p$) is inspired by practical prototyping strategies \cite{RSMA_NOUM_prototype}.}, as they are not obtained by optimizing any performance metrics (e.g., sum throughput for communications, SNR of radar return for sensing). However, they are still useful because they are easy to implement, which in turn makes it relatively easier to obtain an \textbf{empirical} \textbf{ISAC performance region} (e.g., the achievable pairs of sum throughput and radar SNR -- see Definition~\ref{def:isac_perf_region} in page 6 for a precise definition) to judge the communications/sensing trade-off associated with each design choice (i.e., RSMA/SDMA with/without dedicated signal). Nevertheless, the suboptimality of the precoders means that the gap between the different ISAC performance region boundaries could potentially be enlarged through optimized precoders. This is left for future work.
\end{nrem}

%Table~\ref{tab:notation} summarizes the notation used so far. 
Next, we focus on some special cases of (\ref{eq:pc})-(\ref{eq:pr}) that are of interest.

\subsection{Special Cases}
\label{subsec:splcases}
\subsubsection{RSMA ISAC without dedicated sensing signal -- the general case}
\label{3-precoder RSMA (general)}
A key feature of (\ref{eq:pc})-(\ref{eq:pr}) is the fact that communications precoders can be used for sensing, but not vice-versa, as $s_r[\cdot]$ offers no communications benefit. Hence, it is reasonable to question whether the allocated power to a dedicated sensing signal/precoder could be diverted to communications to significantly enhance communications performance, with minimal loss in sensing performance? To investigate this, $t_{\rm comms}$ can be set to 1 on (\ref{eq:pc})-(\ref{eq:pr}) to  yield an RSMA ISAC transmit signal involving three precoders, where each precoder is beneficial for both communications and sensing. 

\subsubsection{RSMA ISAC without dedicated sensing signal -- a soft sensing/communications separation}
\label{3-precoder RSMA (inverted)}
While all precoders are beneficial for both communications and sensing in the previous special case, a balance can be struck wherein some of them prioritize sensing more than communications, while the others achieve the opposite effect. This can be realized, for instance, by setting $\alpha_c = 1 - \alpha_p$, where $\alpha_p \in [1/2, 1)$ (in addition to $t_{\rm comms} = 1$). Thus, the private stream precoders prioritize communications, while the common stream precoder prioritizes sensing, similar to \cite{loli2022ratesplitting}. In this way, a soft separation of communications/sensing functionality for the precoders can be realized.

\subsubsection{SDMA ISAC with dedicated sensing signal -- the general case}
\label{3-precoder SDMA (general)}
With a dedicated sensing signal, $t_{\rm comms} < 1$. Furthermore, from Remark~\ref{rem:sdma_rsma}, it follows that SDMA can be realized by setting $t_p = 1$.

\subsubsection{SDMA ISAC with dedicated sensing signal -- a hard sensing/communications separation}
\label{3-precoder SDMA (separate)}
With a dedicated sensing signal/precoder, perhaps the communications precoder should \emph{solely} prioritize communications to significantly enhance communications performance, with minimal loss in sensing performance? This amounts to a \emph{hard} communications/sensing separation among the precoders, in contrast to the soft separation from special case no. 2. This can be realized by setting $\alpha_p = 1$ (in addition to $t_{\rm comms} <1, t_p = 1$).

\subsubsection{SDMA ISAC without dedicated sensing signal}
\label{2-precoder SDMA}
Finally, this special case can be realized by setting $t_{\rm comms} = 1$ and $t_p = 1$, giving rise to an SDMA ISAC transmit signal involving two precoders, each of which is beneficial for both communications and sensing.

Table~\ref{tab:spl_cases} summarizes the special cases described above\footnote{NOMA is a subset of RSMA obtained by switching off one private stream \cite{RSMA_JSAC_Primer}. However, NOMA ISAC is not evaluated in this paper since it has been shown to provide the lowest ISAC performance in the three-way cross-comparison with SDMA ISAC and RSMA ISAC theoretically \cite{loli2022ratesplitting,YinLF_CommL_RSMA_ISAC_wo}, and lower throughput performance experimentally \cite{lyu2023prototype}.}. Next, we introduce the performance metrics for communications and sensing performance and define the ISAC performance region to quantify ISAC performance.

\begin{table*}[]
    \centering
    \begin{tabular}{|p{2cm}|p{2cm}|c|p{5cm}|}
    \hline
     Scheme & Parameters & Precoders & Comments \\
    \hline
     Spl. 1: RSMA ISAC w/o dedicated sensing signal -- the general case  & \begin{tabular}{c}
        $t_{\rm comms} = 1$ \\
        $\alpha_c\in [0,1]$ \\
        $\alpha_p \in [0,1]$ \\
        $t_p \in [0,1]$ 
    \end{tabular} & \begin{tabular}{c}
        $\nbp_c[k] = \sqrt{P_T (1-t_p)} \times \frac{\sqrt{\alpha_c} \nbu_{c}[k] + \sqrt{1 - \alpha_c} \nbu_0}{\sqrt{\sum\limits_{k'} \|\sqrt{\alpha_c} \nbu_{c}[k'] + \sqrt{1 - \alpha_c} \nbu_0\|^2}}$ \\
        $\nbp_1[k] = \sqrt{P_T \frac{t_p}{2}} \times \frac{\sqrt{\alpha_p} \nbu_1[k] + \sqrt{1 - \alpha_p} \nbu_0}{\sqrt{\sum\limits_{k'} \|\sqrt{\alpha_p} \nbu_1[k'] + \sqrt{1 - \alpha_p} \nbu_0\|^2}} $ \\
        $\nbp_2[k] = \sqrt{P_T \frac{t_p}{2}} \times \frac{\sqrt{\alpha_p} \nbu_2[k] + \sqrt{1 - \alpha_p} \nbu_0}{\sqrt{\sum\limits_{k'} \|\sqrt{\alpha_p} \nbu_2[k'] + \sqrt{1 - \alpha_p} \nbu_0\|^2}} $
    \end{tabular} & $\nbu_c[k] = (\nbu_1[k] + \nbu_2[k])/\|\nbu_1[k] + \nbu_2[k] \|$. See Section~\ref{3-precoder RSMA (general)}. No dedicated radar signal/precoder. In general, common and pvt. streams have different priorities for communications/sensing (i.e., $\alpha_c \neq \alpha_p$).\\
    \hline
     Spl. 2: RSMA ISAC w/o dedicated sensing signal -- a soft sensing/communications separation & \begin{tabular}{c}
        $t_{\rm comms} = 1$ \\
        $\alpha_c = 1 - \alpha_p$ \\
        $\alpha_p \in [1/2, 1]$ \\
        $t_p \in [0,1]$
    \end{tabular}& 
    \begin{tabular}{c}
        $\nbp_c[k] = \sqrt{P_T (1-t_p)} \times \frac{\sqrt{1 - \alpha_p} \nbu_{c}[k] + \sqrt{\alpha_p} \nbu_0}{\sqrt{\sum\limits_{k'} \|\sqrt{1 - \alpha_p} \nbu_{c}[k'] + \sqrt{\alpha_p} \nbu_0\|^2}}$ \\
        $\nbp_1[k] = \sqrt{P_T \frac{t_p}{2}} \times \frac{\sqrt{\alpha_p} \nbu_1[k] + \sqrt{1 - \alpha_p} \nbu_0}{\sqrt{\sum\limits_{k'} \|\sqrt{\alpha_p} \nbu_1 + \sqrt{1 - \alpha_p} \nbu_0\|^2}}$ \\
        $\nbp_2[k] = \sqrt{P_T \frac{t_p}{2}} \times \frac{\sqrt{\alpha_p} \nbu_2[k] + \sqrt{1 - \alpha_p} \nbu_0}{\sqrt{\sum\limits_{k'} \|\sqrt{\alpha_p} \nbu_2[k'] + \sqrt{1 - \alpha_p} \nbu_0\|^2}} $
    \end{tabular}& See Section~\ref{3-precoder RSMA (inverted)}. A special case of 1 above, where a soft sensing/communications separation is realized through the common (private) stream precoder(s) prioritizing sensing (communications). \\
    \hline
    Spl. 3: SDMA ISAC with dedicated sensing signal -- the general case & \begin{tabular}{c}
        $t_{\rm comms}  \in (0,1) $ \\
        $t_p=1 $ \\
        $\alpha_p \in (0,1)$
    \end{tabular}
    & \begin{tabular}{c}
        $\nbp_1[k] = \sqrt{P_T\frac{t_{\rm comms}}{2}} \times \frac{\sqrt{\alpha_p} \nbu_1[k] + \sqrt{1 - \alpha_p} \nbu_0}{\sqrt{\sum\limits_{k'} \|\sqrt{\alpha_p} \nbu_1[k'] + \sqrt{1 - \alpha_p} \nbu_0\|^2}} $ \\
        $\nbp_2[k] = \sqrt{P_T \frac{t_{\rm comms}}{2}} \times \frac{\sqrt{\alpha_p} \nbu_2[k] + \sqrt{1 - \alpha_p} \nbu_0}{\sqrt{\sum\limits_{k'} \|\sqrt{\alpha_p} \nbu_2[k'] + \sqrt{1 - \alpha_p} \nbu_0\|^2}} $ \\
        $\nbp_r[k] = \sqrt{P_T \frac{(1 - t_{\rm comms)}}{N_c}} \times \nbu_0$
    \end{tabular} & See Section~\ref{3-precoder SDMA (general)}. There exists a dedicated radar signal/precoder, but the communications precoders also have some priority for sensing. \\
    \hline
    Spl. 4: SDMA ISAC with dedicated sensing signal -- a hard sensing/communications separation & \begin{tabular}{c}
        $t_{\rm comms} \in  (0,1)$ \\
        $t_p = 1$ \\
        $\alpha_p = 1$
    \end{tabular}
    & \begin{tabular}{c}
        $\nbp_1[k] = \sqrt{P_T \frac{t_{\rm comms}}{2}} \times  \nbu_1[k] $\\
        $\nbp_2[k] = \sqrt{P_T \frac{t_{\rm comms}}{2}} \times  \nbu_2[k] $ \\
        $\nbp_r[k] = \sqrt{P_T \frac{(1 - t_{\rm comms})}{N_c}} \times \nbu_0$
    \end{tabular} & See Section~\ref{3-precoder SDMA (separate)}. A special case of 3 above, where the communications precoders do not prioritize sensing. Thus, a hard sensing/communications separation is realized. \\
    \hline
    Spl. 5: SDMA ISAC w/o dedicated sensing signal & \begin{tabular}{c}
        $t_{\rm comms} =  1$ \\
        $t_p = 1$ \\
        $\alpha_p \in (0,1)$
    \end{tabular}
    & \begin{tabular}{c}
        $\nbp_1[k] = \sqrt{\frac{P_T}{2}} \times \frac{\sqrt{\alpha_p} \nbu_1[k] + \sqrt{1 - \alpha_p} \nbu_0}{\sqrt{\sum\limits_{k'} \|\sqrt{\alpha_p} \nbu_1[k'] + \sqrt{1 - \alpha_p} \nbu_0\|^2}} $ \\
        $\nbp_2[k] = \sqrt{\frac{P_T}{2}} \times \frac{\sqrt{\alpha_p} \nbu_2[k] + \sqrt{1 - \alpha_p} \nbu_0}{\sqrt{\sum\limits_{k'} \|\sqrt{\alpha_p} \nbu_2[k'] + \sqrt{1 - \alpha_p} \nbu_0\|^2}} $
    \end{tabular} & See Section~\ref{2-precoder SDMA}. No dedicated radar signal/precoder \\
    \hline
    \end{tabular}
    \caption{List of special cases described in Section~\ref{subsec:splcases}.}
    \label{tab:spl_cases}
\end{table*}

\subsection{Performance Metrics}
\label{sec:ISAC metric}
Let $\nbP := \{\nbp_c[k], \nbp_1[k], \nbp_2[k], \nbp_r[k]: \forall k \}$ denote the collection of precoders.
\paragraph{Communications} 
In our system, the Modulation and Coding Scheme (MCS) determines the spectral efficiency of a data stream by specifying the modulation order and channel coding rate used for transmission. In our measurements, the communications performance metric is the \emph{MCS-limited sum throughput}. An MCS level is characterized by a pair $(m, r)$, where positive integer $m$ denotes the bits per constellation symbol (e.g., 1 for BPSK) and $r \in (0,1]$ denotes the code rate. 

\begin{nrem}[Necessary condition for decoding MCS level $(m,r)$]
\label{rem:mcs}
An OFDM symbol stream over $N_c$ subcarriers encoded using $(m,r)$ can be successfully decoded at a UE \textbf{only if}
\begin{align}
\label{ineq:mcs_snr}
    \frac{1}{N_c} \sum_{k = 0}^{N_c - 1} \log_2(1 + {\sf SINR}[k]) \geq mr
\end{align}
where ${\sf SINR}[k]$ denotes the signal-to-interference-plus-noise ratio of the symbol stream at subcarrier $k$ at the UE. 

The above inequality is a \textbf{necessary, but not sufficient condition} for successfully decoding $(m,r)$. This is because the left hand side represents the spectral efficiency (in bits/s/Hz) that can be achieved through Gaussian encoding, whereas the right hand side is the spectral efficiency achievable using $(m,r)$, which typically involves encoding over the binary field and modulation over discrete (non-Gaussian) constellations. Due to this mismatch, the minimum SNR needed to decode $(m,r)$ is usually higher than that given by (\ref{ineq:mcs_snr}) -- typically modeled as a $< 2{\rm dB}$ penalty \cite{Shannon_bound_adjustment}. 

Successful decoding of $(m,r)$ yields a throughput (measured in bits/s) of $Bmr$, where $B$ denotes the signal bandwidth (after accounting for signaling overheads, such as cyclic prefix).
\end{nrem}

The received signal, $y_i[k]$, at the $i$-th UE given by:
\begin{align}
\label{eq: yi}
 y_i[k] &= \nbh_i^H[k] \nbx[k] + n_i[k] ~~ (i = 1,2) \notag \\
 &= \nbh_i^H[k] \nbp_r[k] s_r[k] +\nbh_i^H[k] \nbp_c[k] s_c[k] \notag \\
 &~ + \nbh_i^H[k] \nbp_1[k] s_1[k] + \nbh_i^H[k] \nbp_2[k] s_2[k] + n_i[k]
\end{align}
We assume $s_r[\cdot]$ is known at the UEs and can be subtracted before decoding $s_c[\cdot]$, $s_1[\cdot]$ and $s_2[\cdot]$. Let $\ncalM = \{(m_c, r_c), (m_1, r_1), (m_2, r_2)\}$ denote the MCS levels chosen for $(s_c[\cdot], s_1[\cdot], s_2[\cdot])$, respectively. After subtracting $s_r[\cdot]$, each UE independently decodes $s_c[\cdot]$, while treating the interference from $s_1[\cdot]$ and $s_2[\cdot]$ as noise. Thus, from Remark~\ref{rem:mcs}, the maximum MCS-limited common stream throughput, denoted by $T_c(\nbP)$, is given by:
\begin{align}
\label{eq:Tc}
    T_c(\nbP) &= \max_{\ncalM \in \nbbM} ~ Bm_c r_c \\
\label{const:mcs_common}
    &\mbox{s.t.}~ \min_i \frac{1}{N_c} \sum\limits_{k=0}^{N_c - 1} \log_2 \left( 1 + {\sf SINR}_{c,i}[k; \nbP] \right) > m_c r_c, 
\end{align}
where (i) $\nbbM$ denotes the collection of permissible MCS levels for the three streams, which is typically pre-determined through standards (see Table~\ref{tab: Mcs table} for the $\nbbM$ used in our measurements), (ii) ${\sf SINR}_{c,i}[k; \nbP]$ denotes the common stream SINR at the $k$-th subcarrier at UE~$i$, which is a function of $\nbP$ as follows:
\begin{align}
\label{eq:SINR_ci}
    {\sf SINR}_{c,i}[k; \nbP] =  \frac{|\nbh_i^H[k] \nbp_c[k]|^2}{\sum_{j = 1}^{2} |\nbh_i^H[k] \nbp_j[k]|^2 + \sigma^2} .
\end{align}
The minimum in (\ref{const:mcs_common}) is due to both UEs needing to decode $s_c[\cdot]$. Furthermore, $T_c(\nbP) = 0$ when (\ref{const:mcs_common}) is not satisfied even for the lowest MCS level.

%achievable rate, $R_c(\nbP)$, for decoding $s_c[\cdot]$ at both UEs is given by
%\begin{align}
%\label{eq:Rc}
%    R_c (\nbP) &= \min_i~ R_{c,i}(\nbP), ~\mbox{where}\\
%R_{c,i} (\nbP) &= \frac{1}{N_c} \sum\limits_{k=0}^{N_c - 1} \log_2 \left( 1 + \frac{|\nbh_i^H[k] \nbp_c[k]|^2}{\sum_{j = 1}^{2} |\nbh_i^H[k] \nbp_j[k]|^2 + \sigma^2}\right), \notag \\
%    & \hspace*{5cm}  (i = 1,2).
%\end{align}
%where $R_{c,i}(\nbP)$ denotes the achievable rate of decoding $s_c[\cdot]$ at UE $i$, and the .

After decoding $s_c[\cdot]$ and subtracting its contribution from $y_i[\cdot]$, UE~$i$ decodes $s_i[\cdot]$ while treating the interference from the other private stream ($s_j[\cdot],~ j \neq i)$ as noise. Similar to $T_c(\nbP)$, the maximum \emph{MCS-limited} private stream throughput at UE $i$, denoted by $T_i(\nbP)$, is given by:
\begin{align}
\label{eq:Ti}
    T_i(\nbP) &:= \max_{\ncalM \in \nbbM} ~ Bm_i r_i \\
\label{const:mcs_pvt}
    &\mbox{s.t.}~ \frac{1}{N_c} \sum\limits_{k=0}^{N_c - 1} \log_2 \left( 1 + {\sf SINR}_{p,i}[k; \nbP] \right) > m_i r_i, 
\end{align} 
where ${\sf SINR}_{p,i}[k; \nbP]$ denotes the private stream SINR at the $k$-th subcarrier for the $i$-th UE, which is a function of $\nbP$ as follows:
\begin{align}
\label{eq:SINR_pi}
    {\sf SINR}_{p,i}[k; \nbP] =  \frac{|\nbh_i[k]^H \nbp_i[k]|^2}{|\nbh_i[k]^H \nbp_j[k]|^2 + \sigma^2} ~ (j \neq i).
\end{align}
From (\ref{eq:Tc})-(\ref{const:mcs_pvt}), the MCS-limited sum throughput, $T_{\rm sum}(\nbP)$, has the following expression:
\begin{align}
\label{eq:Tsum}
T_{\rm sum}(\nbP) = T_c(\nbP) + \mathbbm{1}(T_c(\nbP) > 0) (T_1(\nbP) + T_2(\nbP)).
\end{align}

\begin{nrem}[Impact of Imperfect SIC and Finite MCS levels]
\label{rem:imperfectSIC}
The indicator function in (\ref{eq:Tsum}) captures the impact of imperfect SIC\footnote{Unlike theoretical analyses which typically model imperfect CSI via stochastic error terms (e.g., \cite{loli2022ratesplitting,Rafael_spawc_2021}), our experimental results inherently capture the aggregate effect of all real-world CSI impairments—including estimation error, feedback quantization, and hardware non-linearities. These impairments are not modeled theoretically here but are directly quantified by the observed throughput collapse.} and finite MCS levels -- i.e., if a UE cannot decode the common stream at the lowest MCS level, then it cannot decode its private stream at any MCS level, resulting in zero throughput. If at least one UE experiences such a \textbf{throughput collapse}, then RSMA is not viable (under the constraint of finite MCS levels), and we assume $T_{\rm sum}(\nbP) = 0$. This phenomenon was experimentally observed in \cite{RSMA_NOUM_prototype}.

It is worth noting that this is a marked departure from standard analytical models that assume perfect SIC and infinitely many MCS levels of arbitrary granularity (implicit when Gaussian encoding is assumed). Under these assumptions, the highest achievable RSMA sum throughput is never smaller than the highest achievable SDMA sum throughput \cite{HamdiMISOImperfectCSIT}. However, with finite MCS levels, the SDMA sum throughput can exceed the RSMA sum throughput, especially when the common stream precoder is not allocated enough power to decode the lowest MCS level.
\end{nrem}

\begin{nrem}[SDMA Sum Throughput]
Since there is no common stream in SDMA (see Remark~\ref{rem:sdma_rsma}), the maximum MCS-limited SDMA sum throughput equals $T_1(\nbP) + T_2(\nbP)$, where $\ncalM = \{(m_1,r_1), (m_2, r_2)\}$.
\end{nrem}

\begin{nrem}[Upper bound on measured throughputs]
\label{rem:ub}
 Since (\ref{const:mcs_common}) and (\ref{const:mcs_pvt}) are necessary but not sufficient conditions for decoding non-Gaussian constellations (see Remark~\ref{rem:mcs}), $T_c(\nbP)$, $T_i(\nbP)~(i = 1, 2)$ and $T_{\rm sum }(\nbP)$ are upper bounds for the measured common stream, private stream and sum throughputs, respectively.    
\end{nrem}

\paragraph{Ranging}
In our measurements, performance metric for ranging is the \emph{post-processing radar SNR}. We consider a single-antenna radar receiver (RX) colocated with the TX. For transmit signal $\nbx[k]$ given by (\ref{eq:x}), the signal at the radar RX, denoted by $y_r[k]$, is given by:
\begin{align}
\label{eq: reflect signal}
    y_r[k] &= \beta (\nba_0^H \nbx[k]) \exp\left(j 2\pi \frac{n_0}{N_c}k \right) + n_r[k], 
\end{align}
where $\beta$ denotes the signal attenuation, $n_0$ is the time domain sample delay, and $n_r[k] \sim \ncalC \ncalN (0, \sigma_r^2)$ the RX noise. The maximum likelihood estimate of $n_0$, denoted by $\hat{n}_0$, is given by:
\begin{align}
    \label{eq:rangebin_ML}
    \hat{n}_0 &= \arg \max_{n} \left| Y_r[n] \right|, \\
    \label{eq:rangedft}
    \mbox{where}~ Y_r[n] & = \sum_{k = 0}^{N_c - 1} y_r[k] (\nbx^H [k] \nba_0) \exp \left(-j 2\pi \frac{n}{N_c} k\right). 
\end{align}
In (\ref{eq:rangedft}), $Y_r[n]~(n = 0, \cdots, N_c - 1)$ is the $N_c$-point DFT of $\{y_r[k] (\nbx^H [k] \nba_0): k = 0, \cdots, N_c - 1\}$ delayed by $n$ samples, and (\ref{eq:rangebin_ML}) amounts to identifying the location of the DFT peak.

The Cramér-Rao Bound (CRB) for estimating ${n}_0$ is given by: 
\begin{equation}
    \label{eq:CRB}
    {\sf CRB}(n_0) = \frac{\sigma^2N_c^2}{8\pi^2\beta^2\sum_{k=0}^{N_c-1}k^2|\mathbf{a}_0^H\mathbf{x}[k]|^2}.
\end{equation}

The post-processing radar SNR, denoted by ${\sf SNR}_{\rm rad}(\nbP)$, is defined as follows:
\begin{align}
    \label{eq:radarSNR}
    {\sf SNR}_{\rm rad}(\nbP) &= \frac{|Y_r[\hat{n}_0]|^2}{\frac{1}{N_{c}-1}\sum\limits_{n \neq \hat{n}_0}|Y_r[n]|^2}
\end{align}
The right-hand side of (\ref{eq:radarSNR}) is a function of $\nbP$, since $Y_r[\cdot]$ in (\ref{eq:rangedft}) is a function of $\nbx[k]$, which in turn is a function of $\nbP$ (see (\ref{eq:x})). 
The post-processed radar SNR can also be denoted as: 
\begin{align}
\label{eq:sigma in snr}
    {\sf SNR_{\rm rad}}(\nbP)=\frac{\beta^2(N_c-1)\sum_{k=0}^{N_c-1}|\mathbf{a}_0^H\mathbf{x}[k]|^2}{\sigma^2},
\end{align}
which shows that with the fixed number of subcarriers, noise variance and signal attenuation factor $\beta$, a larger $\sum_{k=0}^{N_c-1}|\mathbf{a}_0^H\mathbf{x}[k]|^2$ results in a smaller ${\sf CRB}(n_0)$, and a larger $\sf SNR_{\rm rad}$, amounting to better radar sensing performance.
\begin{nrem}[Radar sensing performance metric]
    While the $\sf{CRB}$ is commonly used as a theoretical benchmark for unbiased estimators, its application in our experimental setup is limited. Specifically, our ranging task is conducted in a controlled indoor environment with static targets and high SNR, resulting in deterministic, quantized delay estimates with no observable empirical variance.

Furthermore, the $\sf{CRB}$ is the theoretical lower bound on the variance of an unbiased estimator. However, in our case, the measured variance is effectively always zero, not because our estimator is perfect, but because the resolution is limited by the sampling rate (100 MHz). Specifically, the minimum delay increment we can resolve is 10 ns, corresponding to about 1.5 m in range. Therefore, a direct comparison between the measured variance and the $\sf{CRB}$ is not meaningful in this context.

Given these conditions, we adopt ${\sf SNR_{\rm rad}}$ as our primary sensing performance metric. It is directly measurable, sensitive to target detectability, and reflects the quality of the matched filter output, which is tightly linked to ranging accuracy in our setup.
\end{nrem}

%\begin{nrem}[Impact of Randomess of Comms symbols]
%    Particularly exacerbated by SDMA without dedicated symbol precoder.
%\end{nrem}
%\begin{align}
%    w_r[k] = \frac{\nbx^H[k] \mathbf{1}}{|\mathbf{1}^H \nbx[k]|^2 + \sigma_r^2}
%\end{align}

\begin{ndef}[RSMA ISAC performance region]
   \label{def:isac_perf_region}
    The ordered pair $(T_{\rm sum}(\nbP), {\sf SNR}_{\rm rad}(\nbP))$ captures RSMA ISAC performance. Specifically, the four parameters -- $(t_{\rm comms}, t_p, \alpha_c, \alpha_p)$ -- uniquely determine $\nbP$ (see (\ref{eq:pc})-(\ref{eq:pr}) and Remark~\ref{rem:4params}), which in turn determines $(T_{\rm sum}(\nbP), {\sf SNR}_{\rm rad}(\nbP))$. Thus, the set of all feasible values for the four parameters corresponds to a collection of points $(T_{\rm sum}(\nbP), {\sf SNR}_{\rm rad}(\nbP))$, which we define as the RSMA ISAC performance region, whose boundary (Pareto frontier) captures the limits of achievable ISAC performance for the heuristic precoders given by (\ref{eq:pc})-(\ref{eq:pr}) -- see also Remark~\ref{rem:heuristic_precoders}.  
\end{ndef}

\begin{nrem}
\label{rem:boundary}
    Along the lines of Definition~\ref{def:isac_perf_region}, the ISAC performance region can be defined for each of the special cases in Table~\ref{tab:spl_cases}. By definition, these performance regions are contained within the RSMA ISAC performance region defined above. However, the following questions are of particular interest:
    \begin{itemize}
        \item Can the boundary of the RSMA ISAC performance region in Definition~\ref{def:isac_perf_region} be achieved by one or more of the special cases in Table~\ref{tab:spl_cases}?

        \item How do the boundaries of the SDMA ISAC performance regions and RSMA ISAC performance regions relate to one another?
    \end{itemize}
\end{nrem}

Before addressing the questions in the previous remark in Section~\ref{sec: expt results}, we first describe our RSMA ISAC prototype in the following section.

\section{RSMA ISAC Prototype}
\label{sec:rsma isac prototype}
\begin{figure*}
\centering
    \includegraphics[width=0.8\linewidth]{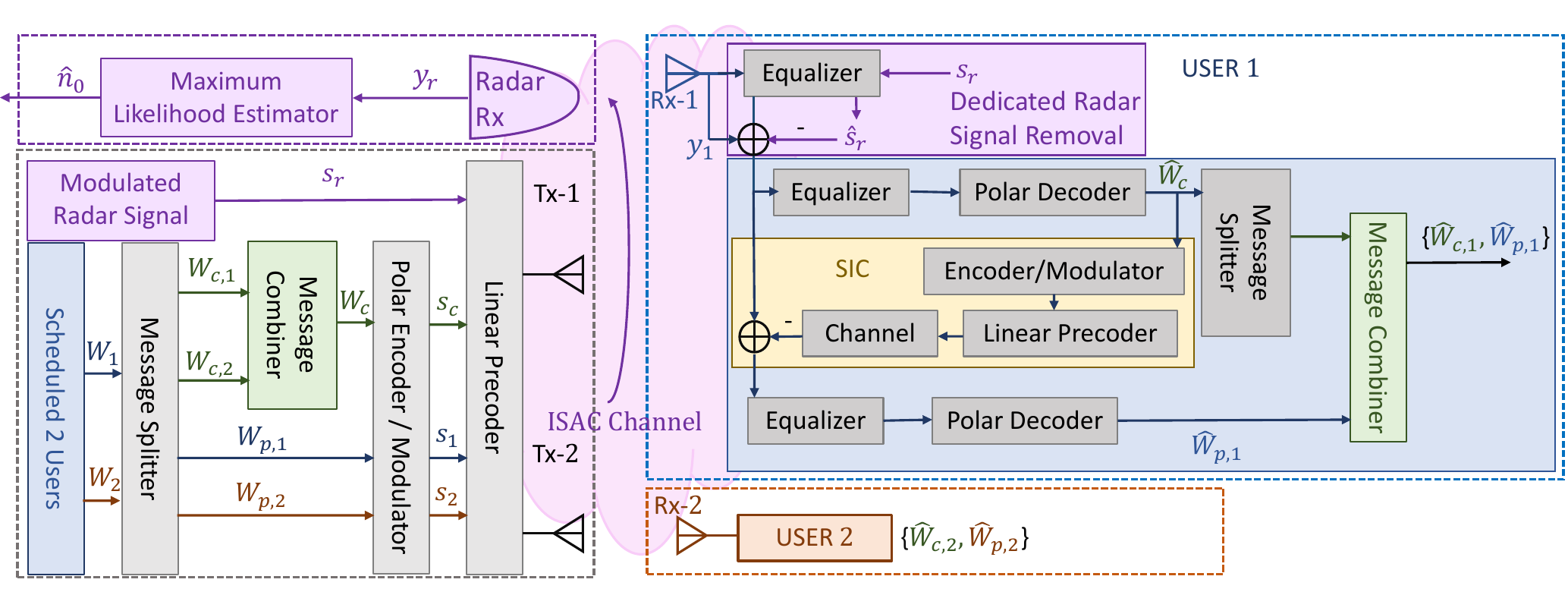}
    \caption{RSMA ISAC block diagram ($N_T = 2$) implemented using SDRs.}
    \label{fig:diagram_rsma_isac}
\end{figure*}
Fig.~\ref{fig:diagram_rsma_isac} depicts the RSMA ISAC block diagram, which we implement using software-defined radios (SDR). The TX, UEs and the radar RX are realized using National Instruments' (NI) USRP 2942 SDR units, which have two antennas/RF chains. Hence, we use three USRP 2942 units -- one to realize the two-antenna TX, another to realize two single-antenna UEs, and a third to realize the (single-antenna) radar RX, where we use a Yagi-antenna pointing towards the target to obtain a strong radar return and suppress some of the self-interference (SI) from the TX $\rightarrow$ radar RX direct path. The TX antenna consists of a patch antenna array\footnote{While the patch antenna array is less directive than a Yagi antenna, it still offers moderate spatial selectivity toward the target region. Unlike omnidirectional antennas (such as monopoles), the patch array provides forward gain and helps suppress side-lobe leakage\cite{balanis2016antenna}, making it a suitable choice for improving signal focus and reducing interference in our experimental ISAC evaluation.} (two-element), which exhibits a directional radiation pattern \cite{balanis2016antenna}. The TX and the radar RX are situated close together to mimic a monostatic configuration. The USRPs share a common timing source (NI CDA-2990), and are controlled by a workstation running LabVIEW NXG, through which the various blocks in Fig.~\ref{fig:diagram_rsma_isac} are realized. All connections (SDRs to workstation, SDRs to timing sources) are through PCIe cables, facilitated by a PCIe bus (NI CPS-8910). A list of hardware components is provided in Table \ref{tab: hardware in RSMA}.
%The radar receiver antenna feeds the received signal to the USRP through a 5m feedback cable. 

\begin{table}
\centering
\begin{tabular}{|l|p{37mm}|p{37mm}|}
\hline
  & \textbf{Name} & \textbf{Description} \\ 
\hline
 1. & Workstation  & Running LabVIEW NXG     \\ 
 2. & NI USRP-2942 (3 units) & SDRs used to realize TX, UEs and radar RX    \\
 3. & NI CPS-8910  & Provides additional PCIe ports\\
 4. & NI CDA-2990 & $8$ Channel, $10{\rm MHz}$ clock distribution device \\ 
 5. & Patch antennas (non-commercial, designed in house) &  TX antennas  \\
 6. & TP-Link TL-ANT2405C (3 units)  & Anchor and (two) UE antennas\\
 7. & MM-ANT-NF-5G Yagi antenna  & Radar RX antenna \\
 \cline{1-3}
\end{tabular}
\caption{List of hardware components in our RSMA prototype.}
\label{tab: hardware in RSMA}
\end{table}

\begin{figure*}
    \centering
    \includegraphics[width=0.75\linewidth]{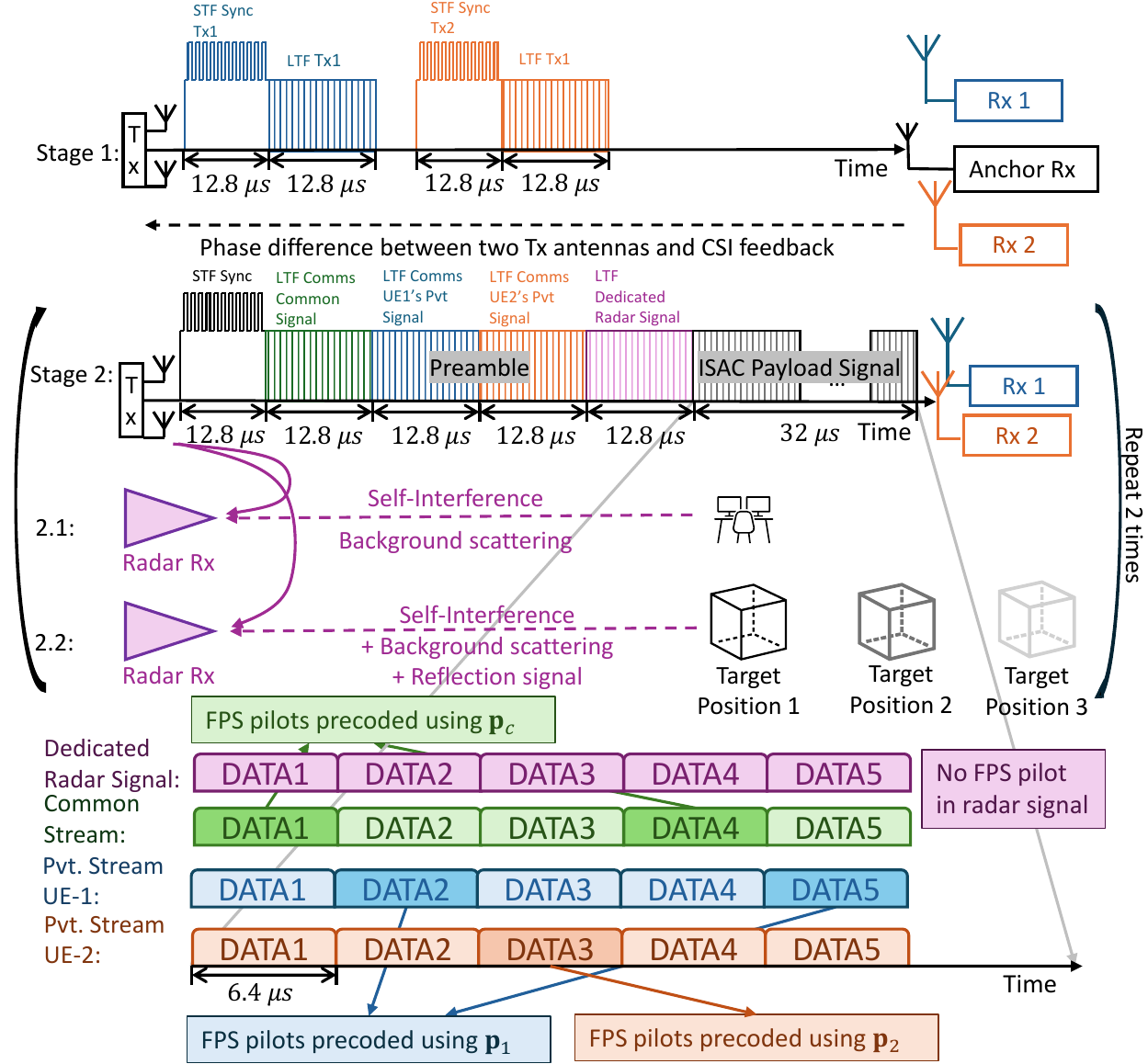}
    \caption{Signal structure within the two-stage transmission protocol used to implement RSMA ISAC in our prototype. In conventional, SDMA-based 802.11ac-VHT, where each UE has to decode only one data stream, every DATA (OFDM) symbol contains 16 \emph{precoded} pilot subcarriers used to correct phase errors in the estimate of a UE's precoded channel. This protects against decoding errors caused by rotated constellations post equalization, and is known as fine phase shift (FPS) \cite{FinephaseShifting}. On the other hand, in RSMA, each UE needs to decode two data streams. To avoid FPS pilot contamination between overlapping streams, we use $\nbp_c$ to precode the FPS pilots in DATA symbols 1 and 4, similarly $\nbp_1$ in DATA symbols 2 and 5, and $\nbp_2$ in DATA symbol 3. The FPS pilots are unused in the stream carrying the dedicated radar signal.}
    \label{fig:ofdm signal stages}
\end{figure*}

We realize the following two-stage protocol to implement the system model described in Section~\ref{sec:sysmodel}, adopting several features of the IEEE 802.11ac-VHT physical layer frames \cite{IEEE_80211_ac}:

\paragraph{Stage 1} 
Each TX antenna transmits a pilot signal orthogonally in time comprising a Short Training Field (STF, $12.8\mu{\rm s}$ in duration) and a Long Training Field (LTF, $12.8\mu{\rm s}$ in duration), as shown in the top portion of Fig.~\ref{fig:ofdm signal stages}. The STF is used for synchronization and coarse frequency offset estimation, while the LTF is used for channel estimation at each UE (i.e., $\hat{\nbh}_i[k])$%{ in Table~\ref{tab:notation}}).

\paragraph{Stage 2} Here, the transmitted signal consists of a preamble followed by the data payload, as shown in the bottom portion of Fig.~\ref{fig:ofdm signal stages}.  
    \begin{itemize}
        \item \textbf{Preamble}: The preamble consists of one STF and four LTFs. The function of the STF is the same as in Stage~1, while the LTFs are precoded in order to estimate the \emph{precoded channels} for equalization at the UEs. The first LTF is used by UE~$i$ to estimate $\nbh_i^H \nbp_c~(i = 1,2)$ for decoding the common stream ($s_c[\cdot]$). The second LTF is used by UE 1 to estimate $\nbh_{1}^H \nbp_1$ to decode its private stream ($s_1[\cdot]$). Similarly, the third LTF is used by UE 2 to estimate $\nbh_{2}^H \nbp_2$ to decode its private stream ($s_2[\cdot]$). The fourth LTF is used by UE $i$ to estimate $\nbh_i^H\nbp_r~(i = 1,2)$ in order to subtract the dedicated sensing signal ($s_r[\cdot])$) from (\ref{eq: yi}).
        
        \item \textbf{ISAC Payload}: For the payload, we consider a total bandwidth of $100{\rm MHz}$ with $N_c = 512$ subcarriers and a cyclic prefix (CP) of $128$ samples per OFDM symbol. Aligned with IEEE 802.11ac-VHT frames, $468$ subcarriers are used to carry data symbols, $16$ subcarriers are used to correct the common phase error across all subcarriers in one OFDM symbol \cite{FinephaseShifting}, and the rest serve as guard bands. This yields an effective bandwidth of:
        \begin{align}
        \label{eq:Beff}
            B &= 100{\rm MHz} \times \underbrace{(512/640)}_\text{CP overhead} \times \underbrace{(468/512)}_{\substack{\text{Guard band} \\ \text{overhead}}} \notag \\
            &= 73.125{\rm MHz}.
        \end{align}
        The ISAC payload consists of at most four superposed streams (one dedicated sensing signal, one common, two private), each comprising $5$ OFDM symbols.

        \item \textbf{MCS Implementation}: Table~\ref{tab: Mcs table} lists the MCS levels, $\nbbM$, implemented in our prototype. For channel coding, we implement Polar codes augmented with an 8-bit cyclic redundancy check \cite{trifonovPolar, constructionPolar}, along with successive cancellation list decoding \cite{listdecoding}, with a list depth of $2$. After the preamble, the first OFDM symbol (labelled SERVICE in the bottom portion of Fig.~\ref{fig:ofdm signal stages}) contains the MCS information of each stream. 
                \begin{table}
        \centering
        \begin{tabular}{|c|c|c|c|}
        \hline
        MCS Index   & Modulation                & Code Rate & Data Rate \\ 
           $\mathcal{M}$   & ($m$)       & $r$       &  $Bmr$ (Mbps)\\
        \hline
        $0$         & BPSK (1)    & $1/2$         & $36.5625$               \\ \hline
        $1$         & BPSK (1)    & $3/4$         & $54.84375$               \\ \hline
        $2$         & QPSK (2)    & $1/2$         & $73.125$              \\ \hline
        $3$         & QPSK (2)    & $3/4$         & $109.6875$              \\ \hline
        $4$         & 16QAM (4)   & $1/2$         & $146.25$              \\ \hline
        $5$         & 16QAM (4)   & $3/4$         & $219.375$              \\ \hline
        $6$         & 64QAM (6)   & $2/3$         & $292.5$              \\ \hline
        $7$         & 64QAM (6)   & $3/4$         & $329.0625$              \\ \hline
        $8$         & 256QAM (8)  & $3/4$         & $438.75$              \\ \hline
        $9$         & 256QAM (8)  & $5/6$         & $487.5$              \\ \hline
        \end{tabular}
        \caption{MCS levels (largely based on IEEE 802.11ac-VHT) implemented in our prototype. The data rate in the last column is equal to $B m r$, where $B$ is the effective bandwidth given by (\ref{eq:Beff}).}
         \label{tab: Mcs table}
        \end{table}

        \item \textbf{Dedicated Sensing Signal}: For $s_r[\cdot]$, we consider a BPSK stream known to the UEs.
        
        \end{itemize}

    % to isolate the target reflections more accurately
    
    % Since the ISAC transceiver operates as a mono-static system and the transmission duration of ISAC system is longer than the sensing signal's round trip time, the radar receiver experiences significant self-interference (SI) from the transmitter. To mitigate this SI effect and background scattering interference on the accuracy of our result, we conduct an initial measurement (Stage 2.1) when no target is placed in front of the ISAC transceiver. This allows us to capture and quantify the SI and background scattering, which are then subtracted from the signals measured in subsequent steps (Stage 2.2-2.4) to isolate the target reflections more accurately\footnote{This method requires sends the same ISAC signal repetitively since SI mitigation and scattering reduction is not this paper's focus, we are using the simplest method to remove it}.

\begin{table}[h]
 \centering
    \begin{tabular}{|l|c|c|}
    \hline 
      Parameter & Notation  & Value\\
    \hline
       Center frequency  & $f_c$  & $2.5{\rm GHz}$ \\
       Transmit power & $P_t$ & $23{\rm dBm}$\\
       Wavelength & $\lambda$ & 0.12m\\
       No. of UEs & & 2 \\
       No. of targets & & 1\\
    \hline   
       Total bandwidth & & $100{\rm MHz}$ \\
       Subcarriers & Total ($N_c$) & $512$ \\
       & Data & $468$ \\
       & Pilot (FPS) & $16$\\
       & Guard band  & $28$ \\
       CP length & & $128$ \\
       Effective bandwidth & $B$ & $73.125{\rm MHz}$ \\
    \hline
       OFDM symbols in payload & & $5$ \\
    \hline
       Range resolution & & $1.5{\rm m}$ \\
       Radar $\rightarrow$ Target distances & & $2.25{\rm m}~(n_0 = 1)$ \\
       & & $3.75{\rm m}~(n_0 = 2)$ \\
       & & $5.25{\rm m}~(n_0 = 3)$ \\
       TX $\rightarrow$ UE distances & & $\approx 1.5{\rm m}$ \\ 
       \hline
       TX antenna spacing & $d$ & $0.0625{\rm m}$ \\
       Fraunhofer distance & & $0.065{\rm m}$\\
       (far-field criterion) & & \\
    \hline 
    \end{tabular}
    \caption{List of miscellaneous parameters in our experiments.}
    \label{tab:param_list}
\end{table}

\section{Experimental Results}
\label{sec: expt results}
\subsection{Measurement Environment}
\label{subsec: meas setup}
\begin{figure}
    \centering
    \begin{subfigure}{0.32\linewidth}
        \centering
        \includegraphics[width = 0.9\linewidth]{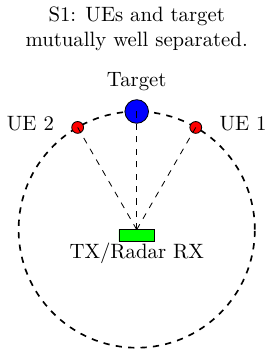}
        \caption{}
        \label{fig:three scenarios a}
    \end{subfigure}%
    \begin{subfigure}{0.32\linewidth}
        \centering
        \includegraphics[width = \linewidth]{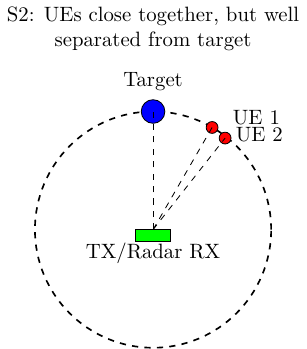}
        \caption{}
        \label{fig:three scenarios b}
    \end{subfigure}%
    \begin{subfigure}{0.32\linewidth}
        \centering
        \includegraphics[width = 0.8\linewidth]{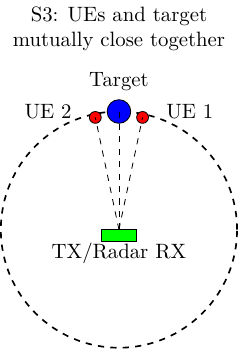}
        \caption{}
        \label{fig:three scenarios c}
    \end{subfigure}
    \par \medskip
    \begin{subfigure}{\linewidth}
        \centering
        \includegraphics[width=\linewidth]{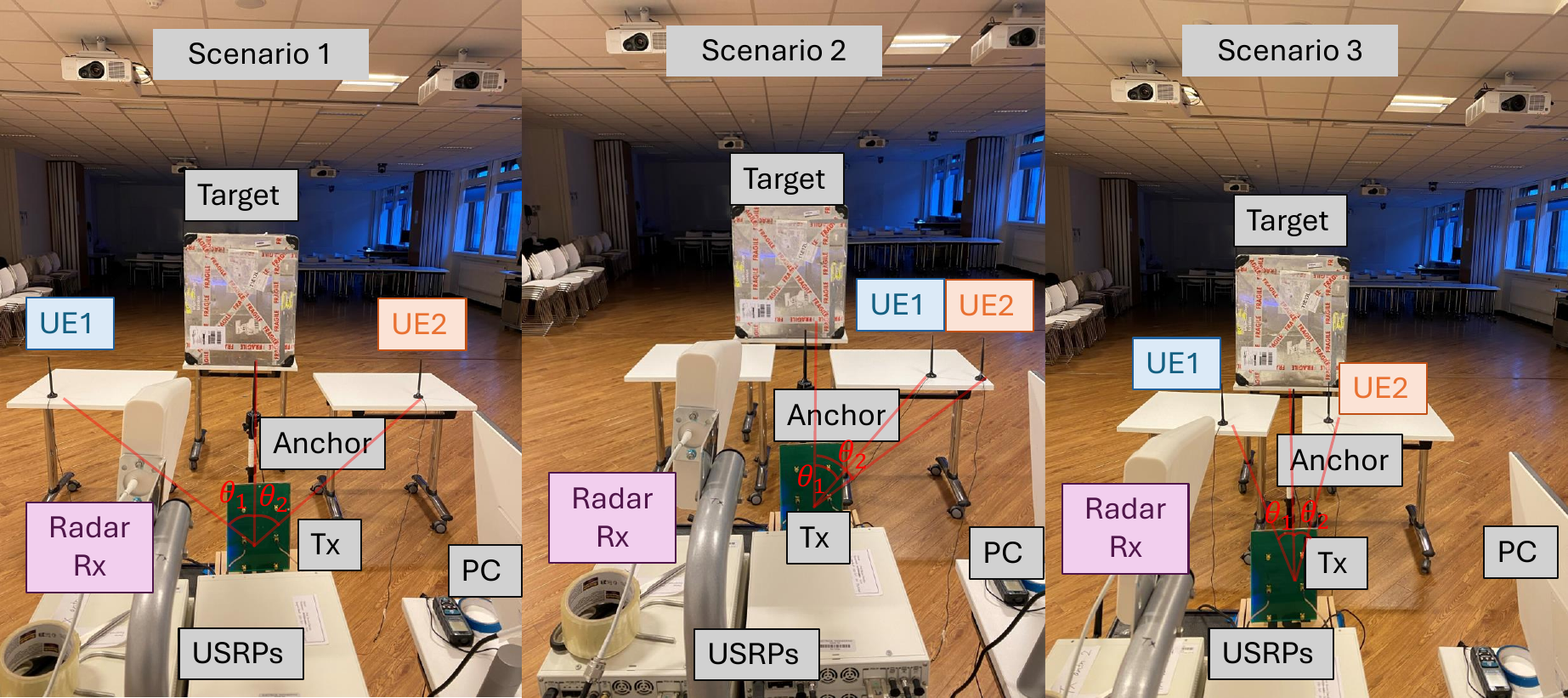}
    \end{subfigure}
    \caption{The three scenarios considered in our measurements.}
    \label{fig:three_setup}
\end{figure}

Our measurements were conducted in a fairly empty lecture room (Fig.~\ref{fig:three_setup}), where the furniture was moved to the far end of the room, well away from the target, to reduce clutter. We used a metal box ($1{\rm m}$ high and $0.8{\rm m}$ wide) placed at the broadside of the TX antennas to mimic a vehicular target. An anchor antenna was also placed at the TX broadside for phase calibration \cite{Radar_conf_2020_doa}. 

We considered three scenarios as described below, capturing different levels of inter-UE interference and separation/integration between communications and sensing:
\begin{itemize}
    \item[S1.] \textbf{UEs and target mutually well separated} (in the angular domain), as shown in {Fig.~\ref{fig:three scenarios a}}. The larger the separation among UEs in the angular domain, the better SDMA can suppress the inter-UE interference. As a result, the RSMA gain -- i.e., the extent to which its ISAC performance region is bigger than SDMA's -- could potentially be modest. Similarly, the extent of separation/integration between communications and sensing separation is dictated by the angular separation between the UEs and the target. A large angular separation may require a dedicated sensing signal to achieve the RSMA ISAC performance boundary.
    
    \item[S2.] \textbf{UEs close together, but well separated from the target}, as shown in {Fig.~\ref{fig:three scenarios b}}. By allocating more power to the common stream, RSMA can better suppress the high inter-UE interference when the UEs are not adequately separated in the angular domain. Hence, RSMA is expected to yield large throughput gains over SDMA in this scenario.

    \item[S3.] \textbf{UEs and targets mutually close together}, as shown in {Fig.~\ref{fig:three scenarios c}}. The high level of integration between communications and sensing and the prominent role played by the common stream in suppressing the inter-UE interference makes this the most likely scenario where a dedicated sensing signal may not be needed for achieving the RSMA ISAC performance boundary. 
\end{itemize}
It is worth pointing out that the geometry of these scenarios are representative of what could be encountered in peer-to-peer vehicular ISAC.

% For communication UEs in scenario 1, both UEs are located on each sides of the target (metal box) with a large angle separation; in scenario 2, both UEs are located on the left side of the target with small angle separation; in scenario 3, both UEs are located closely in the right front of the target. Intuitively, the channel correlation between both UEs are expected to be small in scenario 1 and large in scenarios 2 and 3.

% set-ups for ISAC measurements in three scenarios, in all scenarios the target is located in front of ($0^\circ$) the transmitter antennas. 

% This configuration evaluates radar detection capabilities across different target distances at a $0^\circ$ angle. This figure indicates the UEs positions in scenario 1.

% The figure shows the ISAC measurement set-up with three target positions. The target is positioned at ,  and  from the ISAC transceiver, corresponding to target positions 1, 2 and 3 (in the middle of range bins). 

For each scenario, the goal of our measurements is to empirically characterize the RSMA ISAC performance region as per Definition~\ref{def:isac_perf_region}. However, this can be quite cumbersome with four precoder parameters involved in determining each point $(T_{\rm sum}(\nbP), {\sf SNR}_{\rm rad}(\nbP))$ in the RSMA ISAC performance region; for instance, with each parameter taking values over $[0, 1]$, a granularity of $0.1$ per parameter yields $10^4$ measurements. That said, it is more important to characterize the boundary of the RSMA ISAC performance region than its interior, as the boundary captures the limits of achievable ISAC performance. Thus, Remark~\ref{rem:boundary} motivates the question: can we identify a smaller set of ``well-chosen" precoder parameters corresponding to the boundaries of the RSMA and SDMA ISAC performance regions for a given scenario? We address this question next.

\subsection{Simulation-aided search for precoder parameters}
\label{subsec:sim_param_search}
Given the estimated UE channels ($\hat{\nbh}_i[k]$), one can obtain a \emph{simulated} RSMA ISAC performance region via a fine-grained sweep of the precoder parameters. We make this notion precise through the following definition.

\begin{ndef}[Simulated RSMA ISAC performance region]
\label{def:sim_perf_region}
    The simulated RSMA ISAC performance region is the set of points $(T_{\rm sum}(\nbP), G_0(\nbP))$, where $T_{\rm sum}(\nbP)$ is the \textbf{expected}, rather than measured, sum throughput (see Remark~\ref{rem:ub}), and $G_0(\nbP)$ is the power radiated towards the TX's broadside, given by
        \begin{align}
            G_0(\nbP) = \sum_{k = 0}^{N_c - 1} |\nba_0^H \nbx[k]|^2,
        \end{align}
In Definition~\ref{def:isac_perf_region}, the sensing metric -- ${\sf SNR}_{\rm rad}(\nbP)$ -- is straightforward to evaluate using measurement data (see \ref{eq:radarSNR}). However, to evaluate ${\sf SNR}_{\rm rad}(\nbP)$ in a simulation setting, we need to make several (potentially unrealistic) assumptions regarding the radar return (extent of signal attenuation, clutter etc.) which we wish to avoid. Hence, we use $G_0(\nbP)$, which is proportional to ${\sf SNR}_{\rm rad}(\nbP)$, as the sensing metric for the simulated RSMA ISAC performance region.
\end{ndef}

Ideally, the simulated RSMA ISAC performance region should be a good proxy for the \emph{true} performance region -- in particular, the precoder parameters corresponding to the RSMA/SDMA boundaries should have similar values over both regions; this then yields a smaller set of ``well-chosen" parameter values that can be used for measuring RSMA and SDMA ISAC performance. 

\begin{nrem}[On Simulation Purpose and Practical Role]
    \label{rem:simulation vs practical}
    The simulation results in Fig.~\ref{fig:sim} are intended to reveal the ISAC performance trade-off regions between sensing and communication under different power allocation settings and priority settings. Specifically, the goal is to identify the boundary points and the boundary-achieving parameter configurations that characterize the performance limits of RSMA ISAC and SDMA ISAC. This approach provides guidance on selecting parameter configurations that minimise the number of measurements required in the experimental study.
    
    These results are not designed for direct comparison with radar SNR values, as doing so would require detailed modelling of target reflection loss, system noise sources, and environmental effects, which are difficult to quantify precisely.

Instead, we use the simulation to identify parameter settings that push the system to boundary points (e.g., communication-optimal, sensing-optimal, or trade-off configurations). This approach significantly reduces the experimental workload, as we only need to measure representative configurations that reflect key trade-offs, rather than exhaustively covering the full parameter space.
\end{nrem}

Fig.~\ref{fig:sim} plots the simulated RSMA ISAC region for each scenario by sweeping over the four precoder parameters in increments of $0.1$, resulting in a reasonably comprehensive exploration of the parameter space. The points corresponding to the special cases in Table~\ref{tab:spl_cases} are marked as well. The sum throughput values in these plots are based on \emph{measured} UE channels -- for each scenario, we placed the UEs approximately $1.5{\rm m}$ away from the TX without any target present\footnote{Indeed, the UE channels are affected by the target, which contributes a multipath component (TX $\rightarrow$ Target $\rightarrow$ UE) whose strength depends on the target range. However, for channels of similar strength, as in each of our scenarios, $T_{\rm sum}(\nbP)$ is chiefly determined by the spatial correlation between the channels, which dictates the amount of inter-UE interference. The spatial correlation is not too sensitive to the target range; in fact, it remains qualitatively unchanged (i.e., low in S1, high in S2 and S3) even without a target. Thus, measuring the UE channels without a target present allows for the convenience of generating a single simulated RSMA ISAC performance region per scenario that is an effective proxy for the true performance region, regardless of the target range.}, and estimated the UE channels through only Stage 1 transmission in Fig.~\ref{fig:ofdm signal stages}. We obtain the following insights from Fig.~\ref{fig:sim}:
\begin{figure*}[ht]
    \begin{subfigure}{0.33\textwidth}
        \includegraphics[width=\linewidth]{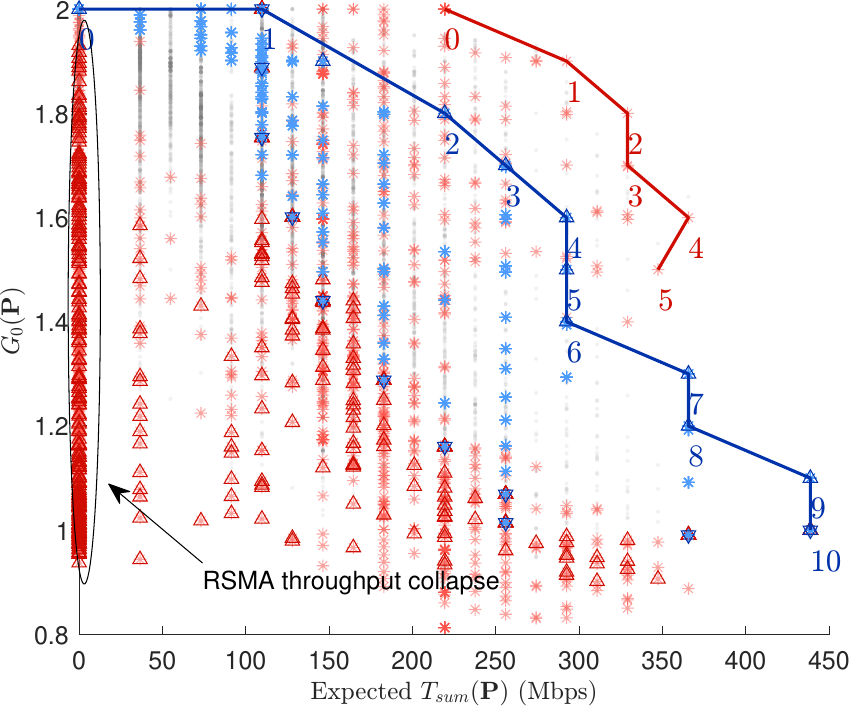}
        \caption{S1 (MRT)}
    \end{subfigure}%
    \begin{subfigure}{0.33\textwidth}
        \includegraphics[width=\linewidth]{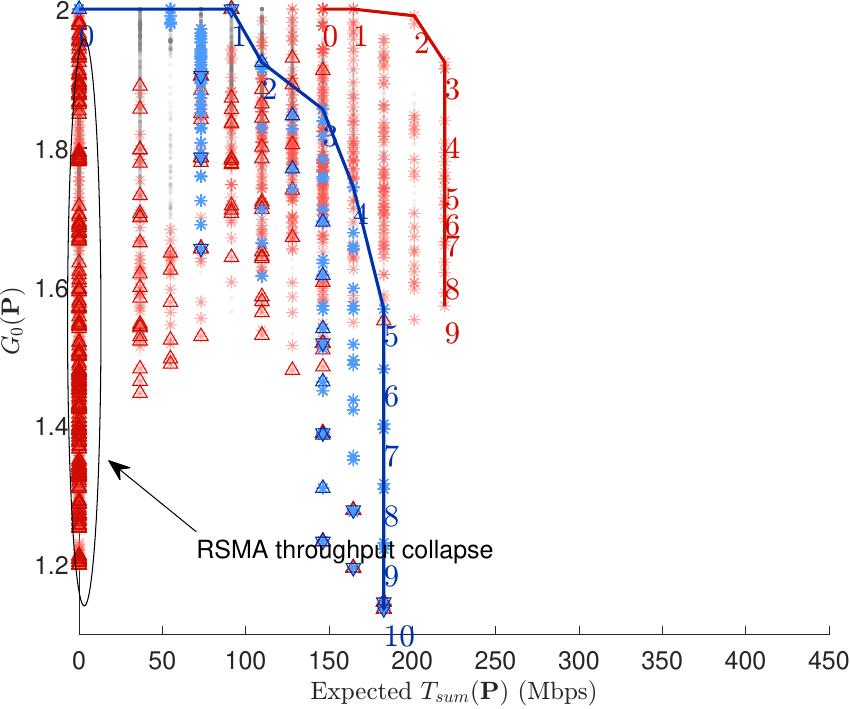}
        \caption{S2 (MRT)}
    \end{subfigure}%
    \begin{subfigure}{0.33\textwidth}
        \includegraphics[width=\linewidth]{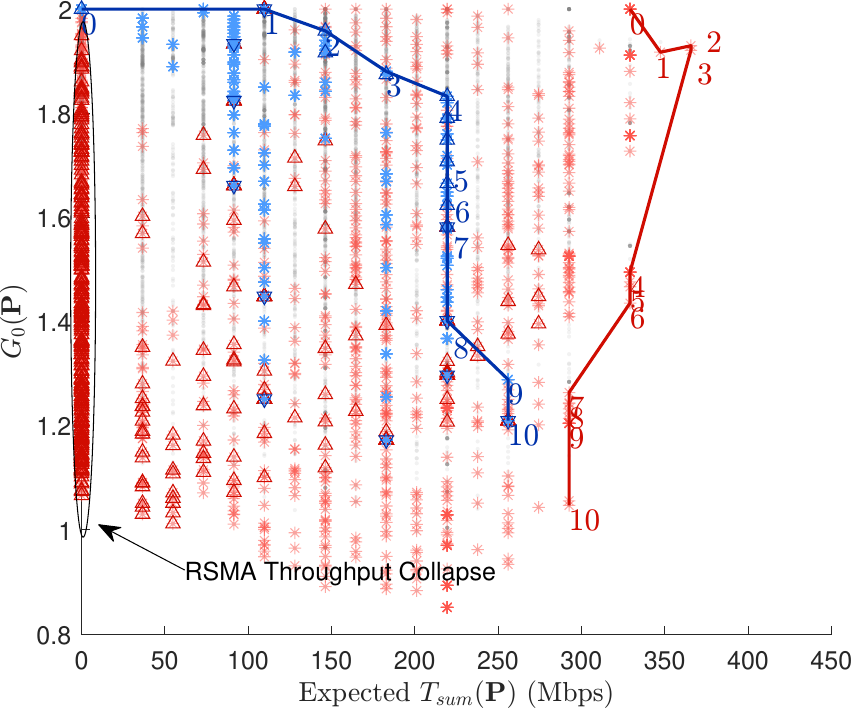}
        \caption{S3 (MRT)}
    \end{subfigure}
    \\
    \begin{subfigure}{0.33\textwidth}
        \includegraphics[width=\linewidth]{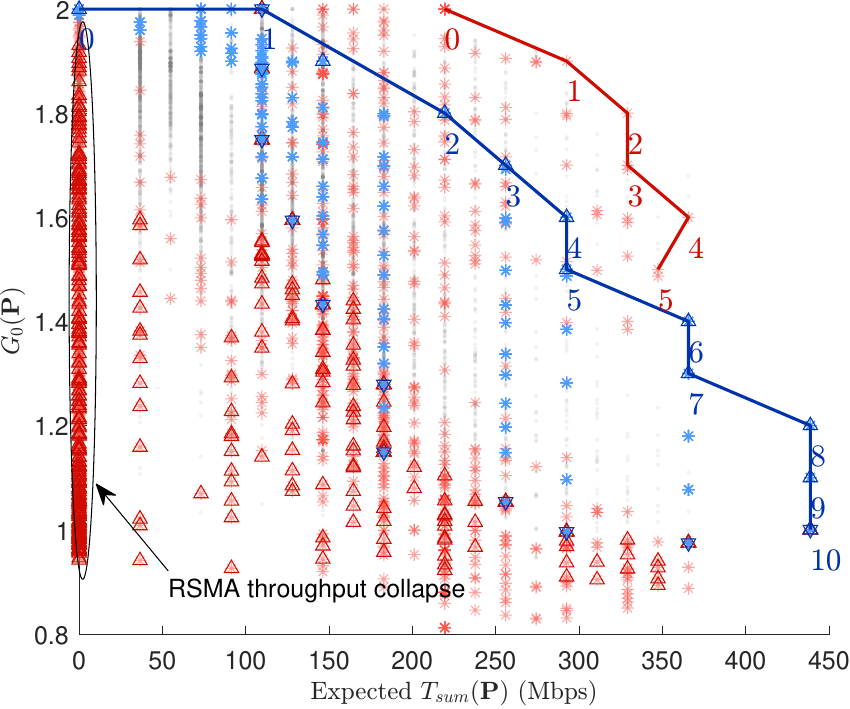}
        \caption{S1 (ZF)}
    \end{subfigure}%
    \begin{subfigure}{0.33\textwidth}
        \includegraphics[width=\linewidth]{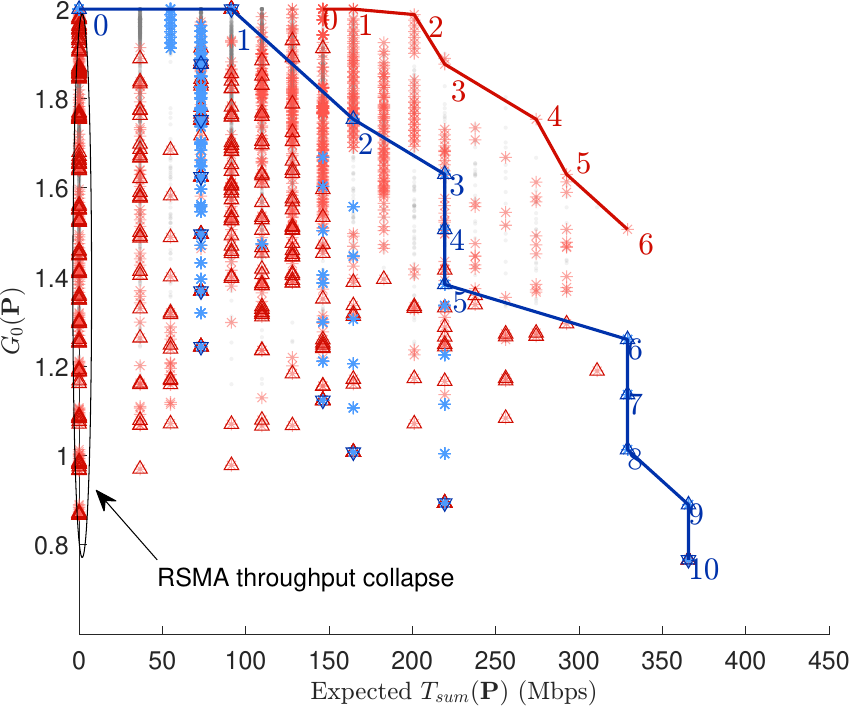}
        \caption{S2 (ZF)}
    \end{subfigure}%
    \begin{subfigure}{0.33\textwidth}
        \includegraphics[width=\linewidth]{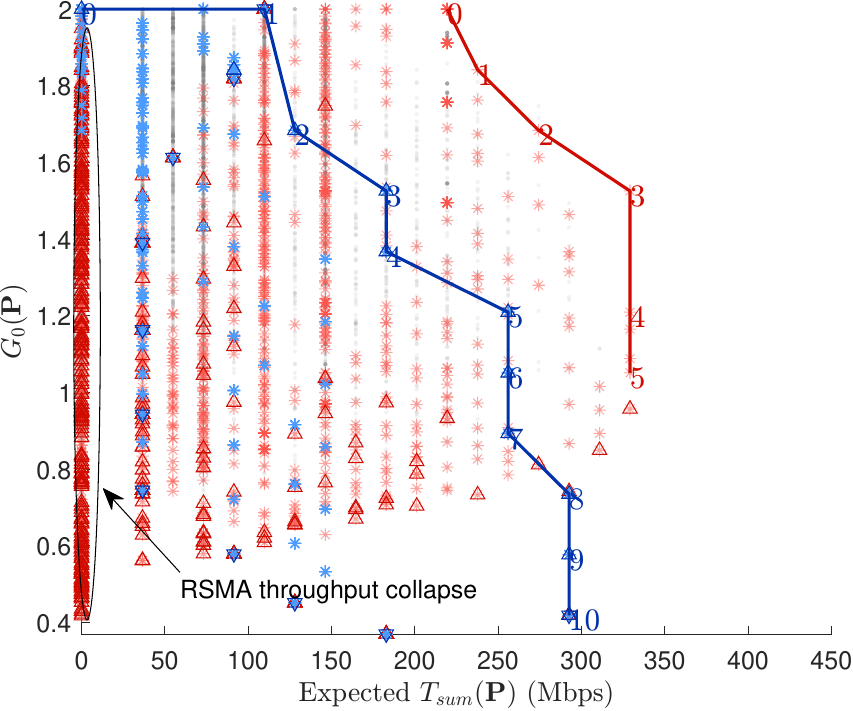}
        \caption{S3 (ZF)}
    \end{subfigure}
    \par \medskip
    \begin{subfigure}{\linewidth}
    \centering
      \footnotesize  
    \begin{tabular}{|c|p{0.42\linewidth}|c|p{0.42\linewidth}|}
    \hline
         \multicolumn{4}{|c|}{Legend}  \\
    \hline
       .  & Most general case: RSMA ISAC with dedicated sensing signal & {\color{blue} $\times$}  & Spl.~Case 3: SDMA ISAC with dedicated sensing signal (general)\\
      {\color{red} $\times$}  & Spl.~Case 1: RSMA ISAC w/o dedicated sensing signal (general) & {\color{blue} $\triangle$}  & Spl.~Case 4: SDMA ISAC with dedicated sensing signal (hard)\\
      {\color{red} $\triangle$}  & Spl.~Case 2: RSMA ISAC w/o dedicated sensing signal (soft)  & {\color{blue} $\Box$}   & Spl.~Case 5: SDMA ISAC w/o dedicated sensing signal \\
     {\color{red} ----}   & RSMA performance boundary & {\color{blue} ----}   & SDMA performance boundary \\
    \hline
    \end{tabular}
    \end{subfigure}
    \caption{Simulated RSMA ISAC performance regions, as per Definition~\ref{def:sim_perf_region}, for each of the scenarios in Fig.~\ref{fig:three_setup}. The parameters associated with the points (marked $0$, 1, etc.) on the SDMA and RSMA performance boundaries are listed in Table~\ref{tab:parameter_both}.}
    \label{fig:sim}
\end{figure*}

\begin{table*}[]
\begin{subtable}{\textwidth}
     \centering
    \begin{tabular}{|c|c|c|c|c|c|c|c|c|c|c|c|c|}
    \hline
        Meas.  &  \multicolumn{4}{c|}{S1} &  \multicolumn{4}{c|}{S2} &  \multicolumn{4}{c|}{S3}\\
        \cline{2-13}
         Index      &  $t_{\rm comms}$  & $\alpha_p$ & $\mathcal{M}_1$ & $\mathcal{M}_2$ &  $t_{\rm comms}$  & $\alpha_p$ & $\mathcal{M}_1$ & $\mathcal{M}_2$ &  $t_{\rm comms}$  & $\alpha_p$ & $\mathcal{M}_1$ & $\mathcal{M}_2$\\ \hline
         0 & 0 & - & -  & -  &  0  &  - & -  & -  &0  &  - & -  & -  \\ \hline
         1 & 0.1  &  1 & 0  & 0&  0.1  &  0.8 & 0  & 0  &0.4  &  0 & 0  & 0  \\ \hline
         2 & 0.2  &  1 & 1  & 1&  0.2  &  0.8 & 0  & 0  &0.1  &  1 & 0  & 0\\ \hline
         3 & 0.3  &  1 & 1  & 2&  0.3  &  0.8 & 0  & 0  &0.2  &  0.9 & 0  & 0\\ \hline
         4 & 0.4  &  1 & 3  & 3&  0.4  &  0.8 & 1  & 1  &0.4  &  1 & 1  & 1\\ \hline
         5 & 0.5  &  1 & 3  & 3&  0.5  &  0.8 & 1  & 1  &0.5  &  0.9 & 1  & 1\\ \hline
         6 & 0.6  &  1 & 3  & 4&  0.6  &  0.8 & 1  & 1  &0.6  &  0.9 & 1  & 1\\ \hline
         7 & 0.7  &  1 & 3  & 4&  0.7  &  0.8 & 1  & 1  &0.9  &  0.7 & 1  & 1\\ \hline
         8 & 0.8  &  1 & 3  & 4&  0.8  &  0.8 & 1  & 1  &1  &  0.7 & 1  & 1\\ \hline
         9 & 0.9  &  1 & 3  & 4&  0.9  &  0.8 & 1  & 1  &1  &  0.9 & 2  & 2\\ \hline
         10 & 1   &  1 & 4  & 4&  1    &  1   & 2  & 2  &1  &  1 & 2  & 2\\ \hline
    \end{tabular}
    \caption{SDMA with MRT precoder design $(t_p = 1)$}
\end{subtable}

\begin{subtable}{\textwidth}
     \centering
    \begin{tabular}{|c|c|c|c|c|c|c|c|c|c|c|c|c|}
    \hline
        Meas.  &  \multicolumn{4}{c|}{S1} &  \multicolumn{4}{c|}{S2} &  \multicolumn{4}{c|}{S3}\\
        \cline{2-13}
         Index      &  $t_{\rm comms}$  & $\alpha_p$ & $\mathcal{M}_1$ & $\mathcal{M}_2$ &  $t_{\rm comms}$  & $\alpha_p$ & $\mathcal{M}_1$ & $\mathcal{M}_2$ &  $t_{\rm comms}$  & $\alpha_p$ & $\mathcal{M}_1$ & $\mathcal{M}_2$\\ \hline
         0 & 0 & - & -  & -  &  0  &  - & -  & -  &0  &  - & -  & -  \\ \hline
         1 & 1  & 0  &  0 & 0  &  1  &  0 & 0  & 0  &1  &  0 & 0  & 0  \\ \hline
         2 & 0.2& 1  &  2 & 2  &  0.2& 1  &  0 & 0    &0.2  &  1 & 0  & 0\\ \hline
         3 & 0.3& 1  &  2 & 2  &  0.3& 1  &  0 & 0    &0.3  &  1 & 0  & 0\\ \hline
         4 & 0.4& 1  &  3 & 3  &  0.4& 1  &  0 & 0    &0.4  &  1 & 1  & 1\\ \hline
         5 & 0.5& 1  &  3 & 3  &  0.5& 1  &  0 & 1    &0.5  &  1 & 1  & 1\\ \hline
         6 & 0.6& 1  &  4 & 4  &  0.6& 1  &  0 & 1    &0.6  &  1 & 1  & 1\\ \hline
         7 & 0.7& 1  &  4 & 4  &  0.7& 1  &  1 & 1    &0.7  &  1 & 3  & 2\\ \hline
         8 & 0.8& 1  &  5 & 5  &  0.8& 1  &  1 & 1    &0.8  &  1 & 3  & 2\\ \hline
         9 & 0.9& 1  &  5 & 5  &  0.9& 1  &  1 & 2    &0.9  &  1 & 3  & 3\\ \hline
         10 & 1 & 1  &  5 & 5  &  1  & 1  &  1 & 2    &1  &  1 & 4  & 3\\ \hline
    \end{tabular}
    \caption{SDMA with ZF precoder design $(t_p = 1)$}
\end{subtable}

\par \bigskip
\begin{subtable}{\textwidth}
     \centering
    \begin{tabular}{|c|c|c|c|c|c|c|c|c|c|c|c|c|c|c|c|c|c|c|}
    \hline
        Meas.  &  \multicolumn{6}{c|}{S1} &  \multicolumn{6}{c|}{S2} &  \multicolumn{6}{c|}{S3}\\
        \cline{2-19}
         Index      &  $t_{p}$  & $\alpha_c$ & $\alpha_p$ &$\mathcal{M}_c$ &$\mathcal{M}_1$ & $\mathcal{M}_2$ &  $t_{p}$  & $\alpha_c$ & $\alpha_p$ &  $\mathcal{M}_c$ &$\mathcal{M}_1$ & $\mathcal{M}_2$ &  $t_{p}$  & $\alpha_c$ & $\alpha_p$ & $\mathcal{M}_c$& $\mathcal{M}_1$ & $\mathcal{M}_2$\\ \hline
         0 & 0   &  1 &  1 & 5  & -  & -  &  0 &  0 &  0   & 4  & -  & -  &0   &  0 &  0 & 5 &-  & -  \\ \hline
         1 & 0.1 &  1 &  1 & 3  & 0  & 0&  0.2 &  0 &  0.2 & 2  & 0  & 0  &0.1 &  0 &0.6 & 4 &0  & 0 \\ \hline
         2 & 0.2 &  1 &  1 & 2  & 1  & 1&  0.2 &  0 &  0.3 & 2  & 0  & 0  &0.1 &  0 &0.8 & 4 &0  & 0\\ \hline
         3 & 0.3 &  1 &  1 & 1  & 3  & 3&  0.2 &  0 &  0.4 & 2  & 0  & 0  &0.1 &  0 &0.7 & 3 &0  & 0\\ \hline
         4 & 0.4 &  1 &  1 & 0  & 3  & 3&  0.2 &  0 &  0.5 & 2  & 0  & 0  & 0  &  1 &  1 & 5 &-  & -\\ \hline
         5 & 0.5 &  1 &  1 & 0  & 3  & 5&  0.2 &  0 &  0.6 & 1  & 0  & 0  &0.3 &  1 &  0.9 & 3 &1  & 1\\ \hline
         6 &  - &  - &   - & -  &  -&  -& 0.2 &  0 &  0.7  & 1  & 1  & 1  &0.3 &  1 &  0.8 & 3 &0  & 0\\ \hline
         7 &  - &  - &   - & -  &  -&  -& 0.2 &  1 &  0.4  & 3  & 0  & 0  &0.3 &0.9 &  0.9 & 3 &1  & 1\\ \hline
         8 &  - &  - &   - & -  &  -&  -& 0.2 &  1 &  0.6  & 2  & 0  & 1  &0.1 &0.9 &  1  & 4 & 0  & 0\\ \hline
         9 &  - &  - &   - & -  &  -&  -& 0.2 &  1 &  0.7  & 3  & 0  & 1  &0 &  0.9 &  1  & 4 &1  & 1\\ \hline
         10 & - &  - &   - & -  &  -&  -& -   &  - &  -   & -   & -  & -  &0.1& 0.8 & 0.7 & 3 &0  & 0\\ \hline
    \end{tabular}
    \caption{RSMA without dedicated sensing signal ($t_{\rm comms} = 1$) and private streams in MRT precoder}
\end{subtable}

\begin{subtable}{\textwidth}
     \centering
    \begin{tabular}{|c|c|c|c|c|c|c|c|c|c|c|c|c|c|c|c|c|c|c|}
    \hline
        Meas.  &  \multicolumn{6}{c|}{S1} &  \multicolumn{6}{c|}{S2} &  \multicolumn{6}{c|}{S3}\\
        \cline{2-19}
         Index      &  $t_{p}$  & $\alpha_c$ & $\alpha_p$ &$\mathcal{M}_c$ &$\mathcal{M}_1$ & $\mathcal{M}_2$ &  $t_{p}$  & $\alpha_c$ & $\alpha_p$ &  $\mathcal{M}_c$ &$\mathcal{M}_1$ & $\mathcal{M}_2$ &  $t_{p}$  & $\alpha_c$ & $\alpha_p$ & $\mathcal{M}_c$& $\mathcal{M}_1$ & $\mathcal{M}_2$\\ \hline
         0 & 0   &  0 &  - & 5  & -  & -  &  0 &  0 &  -   & 5  & -  & -  &0   &  0 &  - & 5 &-  & -  \\ \hline
         1 & 0.1 &  0 &  1 & 3  & 0  & 0&  0.1 &  0 &  0 & 3  & 0  & 0    &0.1 &  0 &  1 & 3 &0  & 0 \\ \hline
         2 & 0.2 &  0 &  1 & 2  & 1  & 1&  0.1 &  0 &  0.1 & 3  & 0  & 0  &0.2 &  0 &  1 & 3 &0  & 0\\ \hline
         3 & 0.3 &  0 &  1 & 1  & 3  & 3&  0.1 &  0 &  1 & 3  & 0  & 0    &0.3 &  0 &  1 & 2 &1  & 0\\ \hline
         4 & 0.4 &  0 &  1 & 1  & 3  & 3&  0.2 &  0 &  1 & 3  & 0  & 0    &0.5  & 0 &  1 & 0 &2 & 2\\ \hline
         5 & 0.5 &  0 &  1 & 0  & 4  & 4&  0.3 &  0 &  1 & 3  & 0  & 0    &0.7 &  0 &  1 & 0 &3  & 3\\ \hline
         6 &  - &  - &   - & -  &  -&  -& 0.3 &  1 &  1  & 3  & 0  & 1    &- &  - &  - & - &-  & -\\ \hline
    \end{tabular}
    \caption{RSMA without dedicated sensing signal ($t_{\rm comms} = 1$) and private streams in ZF precoder}
\end{subtable}
    \caption{Precoder parameters associated with the boundary points of the simulated ISAC performance regions in Fig.~\ref{fig:sim} (the numbers marked $0$, $1$, and so on, in Fig.~\ref{fig:sim} correspond to the measurement index in the first column). The MCS levels shown are the ones that maximize the \emph{expected} sum throughput, based on Remark~\ref{rem:mcs}. These parameters are used for our measurements in Section~\ref{subsec:measured_isac_perf}.}
    \label{tab:parameter_both}
\end{table*}

\paragraph{Impact of finite MCS levels} Perhaps the most striking feature of Fig.~\ref{fig:sim} is the \emph{non-convex} nature of the ISAC performance regions. In particular, the discretized nature of $T_{\rm sum}(\nbP)$ is a direct consequence of assuming finite MCS levels -- instead of a one-to-one mapping between the throughput and SINR (i.e., $\log(1+{\sf SINR})$), a range of SINR values map to the same $T_{\rm sum}(\nbP)$.

\paragraph{Impact of imperfect SIC} A significant proportion of RSMA points lie on the $y$-axis, where $T_{\rm sum}(\nbP) = 0$. These capture the \emph{throughput collapse} phenomenon, where for at least one of the UEs, the SINR is insufficient to decode the common stream at the lowest MCS level in Table~\ref{tab: Mcs table} (see also Remark~\ref{rem:imperfectSIC}). Throughput collapse also explains why the RSMA ISAC performance boundary does not extend as far down as the SDMA boundary in the southeast region of Figs.~\ref{fig:sim}a-b. The common stream SINR reduces as we move from northwest to southeast along the RSMA boundary; when it becomes low enough to trigger a throughput collapse, the boundary ends abruptly. 
    
%    the impact of finite MCS levels introduces two characteristics to the ISAC performance region. As a result, the boundary regions for both SDMA and RSMA are not smooth curves but appear as connected broken lines. For example, in Fig.~\ref{fig:sim}a, points 4, 5 and 6 on the blue curve (SDMA rate/radar region) all exhibit the same sum capacity, but point 4 achieves higher total radar beamforming gain. This suggests that some boundary points are not optimal for expanding the ISAC region, as they do not provide a smooth progression in sum capacity performance. This discontinuity becomes more pronounced in scenarios 2 and 3. Moreover, the effect of limited MCS levels can also be observed in the distribution of RSMA points (red stars) along the y-axis. These points achieve zero capacity because the common stream's capacity, under the given parameter settings, is insufficient to support the lowest MCS level required for successful decoding. Consequently, the imperfect SIC results in residual interference, which contaminates the decoding process of the private streams. This limitation illustrates how finite MCS levels can significantly influence the ISAC regions' shape and characteristics of result distributions, highlighting the importance of considering the MCS constraints in the evaluation of RSMA and SDMA.
\paragraph{Impact of Interference} In scenarios S2 and S3, the peak (expected) sum throughput of SDMA with MRT precoder reduces by more than half relative to S1, due to higher inter-user interference. In contrast, the peak sum throughput of SDMA with ZF precoder reduces by less than half in the same scenarios, which shows ZF can more effectively suppress the inter-user interference in scenarios S2 and S3. By using the ZF precoder, SDMA ISAC achieves a larger ISAC region than using the MRT precoder design.

\paragraph{RSMA ISAC outperforms SDMA ISAC} The SDMA ISAC performance boundary lies in the interior of the RSMA ISAC performance region, thereby addressing the second question posed in Remark~\ref{rem:boundary}. Hence, for well-chosen precoders (which boils down to well-chosen parameters), RSMA ISAC achieves points in the northeast region beyond the SDMA boundary. Furthermore, the extent to which RSMA ISAC outperforms SDMA ISAC is scenario dependent. In scenario S1, where there is a relatively larger angular separation between the UEs, the gap between the RSMA and SDMA boundaries is narrower. The gap becomes much wider -- especially around the northeast corner -- as the UEs move closer together in S2 and S3\footnote{In Fig.~\ref{fig:sim}c, only points 0 to 3 should be considered representative of RSMA's performance boundary for achieving desirable outcomes. The red boundary points from 4 to 10 fall outside the effective performance boundary for RSMA in ISAC evaluation, as their parameter settings cause both $T_{sum}(\nbP)$ and $G_0(\nbP)$ to decrease. We continue to measure these points because they represent the easternmost extent of the red points.}. In Fig.~\ref{fig:sim}d-f, SDMA ISAC with ZF precoder design can achieve a larger ISAC region compared to SDMA ISAC with MRT precoder design, but still RSMA ISAC achieves points in the northeast region beyond the SDMA boundary.

%    In scenario 1, uses have a large angle separation; the performance for both RSMA and SDMA tends to be better, as there is less interference between users, and RSMA and SDMA both achieve larger sum capacity compared to the other two cases. 
    
%    As the users move closer (compared to scenario 1), the performance gap between RSMA and SDMA becomes more evident. In scenarios S2 and S3, RSMA shows higher resilience in interference management than SDMA, as indicated by its higher beamforming gain at the same MCS-limited sum capacity. This shows that RSMA can better balance the trade-off between communication capacity and radar power under tighter user spacing scenarios.

    \paragraph{RSMA ISAC performance boundary achieved without a dedicated sensing signal} In each scenario, the points on the RSMA ISAC performance boundary are achieved when $t_{\rm comms} = 1$, which corresponds to special case 1 in Table~\ref{tab:spl_cases} where the entire transmit power is allocated for communications. This insight addresses the first question posed in Remark~\ref{rem:boundary}.
    
 % varying the beamforming prioritization parameter ($\alpha_c$ and $\alpha_p$) and the power fractions allocated to the private stream ($t_p$). 

    \paragraph{Dedicated sensing signal needed to achieve SDMA ISAC performance boundary} Unlike in d), the points corresponding to SDMA ISAC without a dedicated sensing signal (i.e., special case 5 in Table~\ref{tab:spl_cases}) do not lie on the SDMA ISAC performance boundary.
        
%    For SDMA, the boundary points can be obtained by gradually adjusting the power allocation from one extreme (communication-only, $t_{\rm comms}=1$) to the other (radar-only, $t_{\rm comms}=0$). 
    
Next, we seek to validate the above insights through measurements by selecting precoder parameters listed in Table~\ref{tab:parameter_both} associated with the SDMA and RSMA ISAC boundary points marked $0$, $1$, etc., in Fig.~\ref{fig:sim}.
    
%    We get a lot of insight regarding the questions posed in Remark~\ref{rem:boundary}

%    This insight also helps us to reduce the practical experimental workload to determine the largest achievable region for both schemes; we don't need to do an exhaustive search across all parameter settings.
    
% In the following subsection, we  via measurements.

\subsection{Measurements}
\label{subsec:measured_isac_perf}
% How many measurements

We conducted a total of $336$ measurements\footnote{The ZF- and MRT-based experiments were conducted at different times, and as such, absolute SNR conditions may vary slightly due to changes in the indoor environment (e.g., temperature, multipath, noise floor). However, since the main goal is to compare the performance trends and trade-off boundaries, this variation does not affect the validity of the results.} with the following breakdown:
\begin{itemize}
    \item RSMA -- $138$ measurements ($6$ boundary points for S1-MRT, $6$ boundary points for S1-ZF, $10$ for S2-MRT, $7$ for S2-ZF, $11$ for S3-MRT and $6$ for S3-ZF, along with 3 different target ranges per boundary point to evaluate the effectiveness of ranging, especially for RSMA without a dedicated sensing signal; more details in paragraph a) below).
        
    \item SDMA -- $198$ measurements ($11$ boundary points each for S1, S2 and S3 with both ZF and MRT precoder design, along with 3 different target ranges per boundary point).
\end{itemize}
In these measurements, the UE locations were unchanged from the ones used to obtain Fig.~\ref{fig:sim}. To suppress the radar self-interference (i.e., signal propagation along the direct path from the TX to the radar RX) and background clutter, we perform a simple background subtraction, where Stage~2 is repeated twice per measurement (illustrated as Stages 2.1 and 2.2 in Fig.~\ref{fig:ofdm signal stages}). In the first instance, we obtain the radar return without any target present (Stage 2.1) to capture only the radar self-interference and any background clutter. This signal is then subtracted from the radar return obtained when the target is present (Stage 2.2).

\begin{nrem}[Two-stage self-interference cancellation]
\label{rem:two stage SI cancellation}
    The self-interference cancellation method used in this experiment requires the TX to repeat Stage 2 twice (transmit the same waveform twice). While this repeated transmission is not feasible in practical ISAC systems, it is a controlled lab strategy that allows for precise measurement of the radar reflection signal. This setup allows us to isolate and measure the effectiveness of RSMA ISAC in terms of reflection detection and suppress the self-interference effectively. In practice, real-time self-interference mitigation could be achieved through combinations of antenna isolation, analog cancellation, digital suppression, and adaptive filtering \cite{Liu_erzi_ISAC_SI_JSAC_2023}. Our use of a directional Yagi antenna complements this setup by enhancing the radar return and attenuating self-interference.
\end{nrem}

 Table~\ref{tab:parameter_both} lists the miscellaneous parameters used in our measurements.

% In throughput measurement, we measured the maximum achievable sum throughput by searching the neighbours of the MCS-level in the simulation results, which is based on (\ref{eq:Tsum}). This method can guarantee we can get the MCS level for each stream that maximizes the achievable sum throughput despite the differences between simulation results and practical measurements.

\begin{figure}
    \centering
    \includegraphics[width=\linewidth]{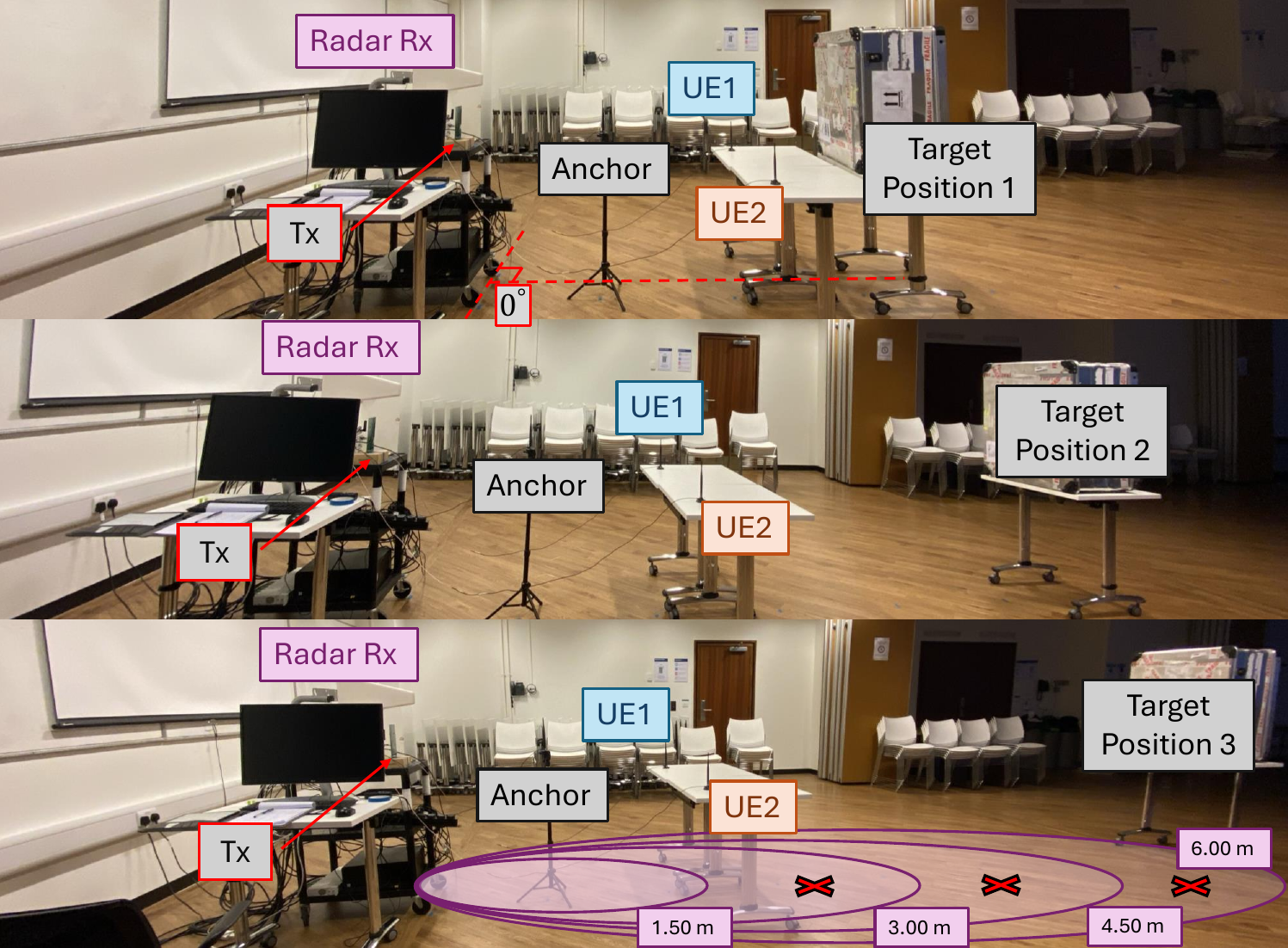}
    \caption{The three target distances considered for each scenario in Fig.~\ref{fig:three_setup} -- $2.25{\rm m}$ (top), $3.75{\rm m}$ (middle) and $5.25{\rm m}$ (bottom), which corresponds to the middle of the second ($n_0 = 1$), third ($n_0 = 2$) and fourth ($n_0 = 3$) range bins, respectively. The $100{\rm MHz}$ signal bandwidth yields a range resolution (bin size) of $1.5{\rm m}$, which is marked in the bottom figure.}
    \label{fig:rangebin}
\end{figure}

\label{sec:isac_performance}
\paragraph{Ranging performance of RSMA ISAC without a dedicated sensing signal}
The $100{\rm MHz}$ signal bandwidth yields a range resolution of $1.5{\rm m}$. Hence, in our measurements, we considered target ranges of $2.25{\rm m}~(n_0 = 1)$, $3.75{\rm m}~(n_0 = 2)$ and $5.25{\rm m}~(n_0 = 3)$ for each scenario, corresponding to the middle of the second, third and fourth range bins, respectively\footnote{$n_0 = 0$ corresponds to the first range bin.}, as shown in Fig.~\ref{fig:rangebin}.

\begin{figure}
    \includegraphics[width=\linewidth]{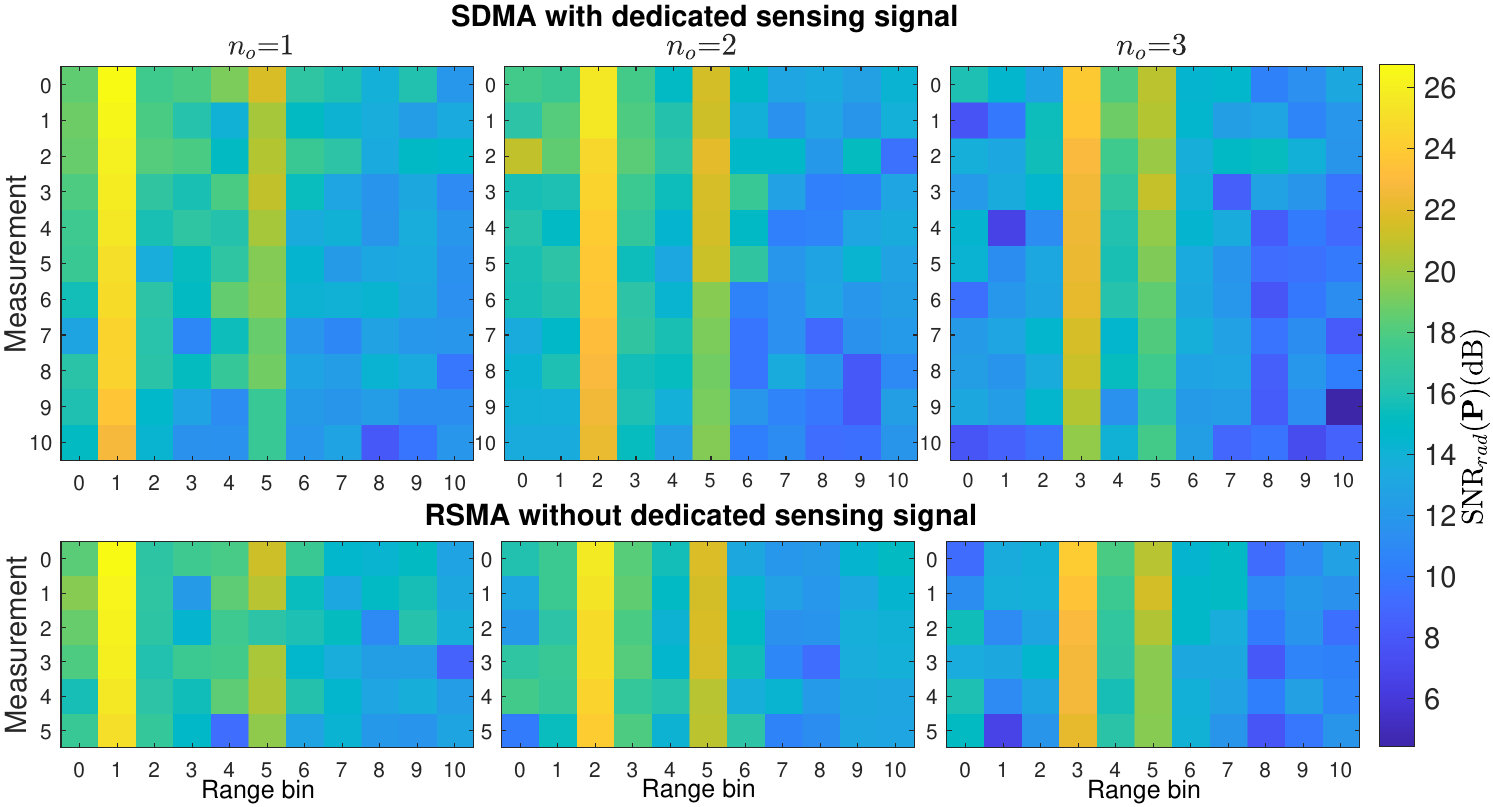}
    \caption{Heatmap plotting the measured ${\sf SNR}_{\rm rad}(\nbP)$ for scenario S1 with MRT precoder design across three different target distances -- $2.25{\rm m}~(n_0 = 1)$, $3.75{\rm m}~(n_0 = 2)$ and $5.25{\rm m}~(n_0 = 3)$. (Top row): SDMA with a dedicated sensing signal; (bottom row) RSMA without a dedicated sensing signal. The $y$-axis in each heatmap corresponds to the measurement index in Table~\ref{tab:parameter_both}. RSMA without a dedicated sensing signal achieves a ranging performance comparable to SDMA with a dedicated sensing signal.}
    \label{fig:ranging_heatmaps}
\end{figure}

For scenario S1, Fig.~\ref{fig:ranging_heatmaps} compares the ranging performance of RSMA (without a dedicated sensing signal) and SDMA (with a dedicated sensing signal). The heatmaps plot ${\sf SNR}_{\rm rad}(\nbP)$ for each range bin. In each heatmap, the target is clearly identifiable at the correct range bin. Moreover, there is a graceful degradation in ${\sf SNR}_{\rm rad}(\nbP)$ with increasing $n_0$ (left to right along each row), consistent with a progressively weaker radar return. Importantly, the absence of a dedicated sensing signal for RSMA does not adversely impact its ranging performance. The heatmaps for scenarios S2 and S3 look similar and hence, not presented.

\paragraph{Measured ISAC performance region -- RSMA v/s SDMA}
\begin{figure*}
    \centering % <-- added
\begin{subfigure}{0.3\textwidth}
  \includegraphics[width=\linewidth]{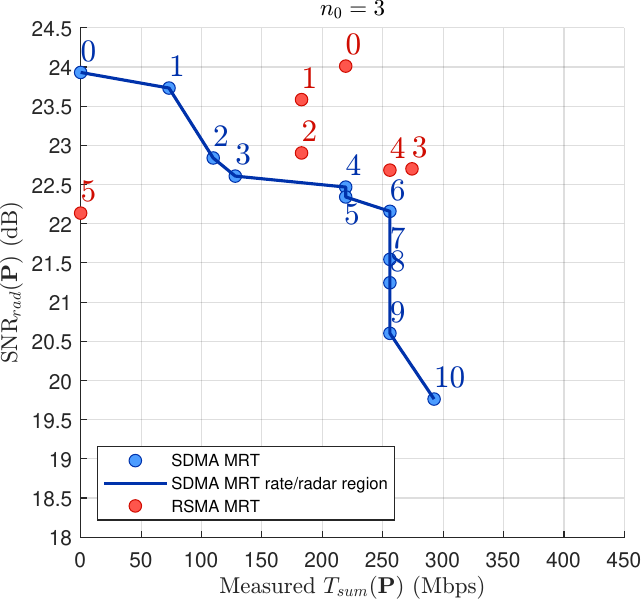}
  \caption{S1 (MRT)}
\end{subfigure}\hfil % <-- added
\begin{subfigure}{0.3\textwidth}
  \includegraphics[width=\linewidth]{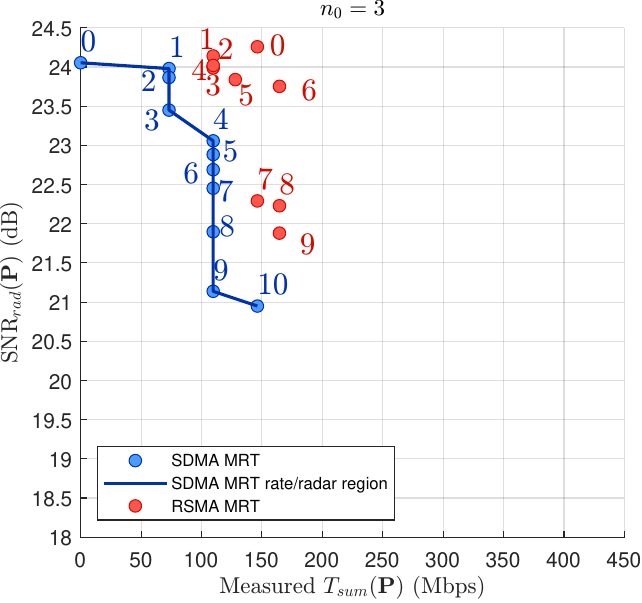}
  \caption{S2 (MRT)}
\end{subfigure}\hfil % <-- added
\begin{subfigure}{0.3\textwidth}
  \includegraphics[width=\linewidth]{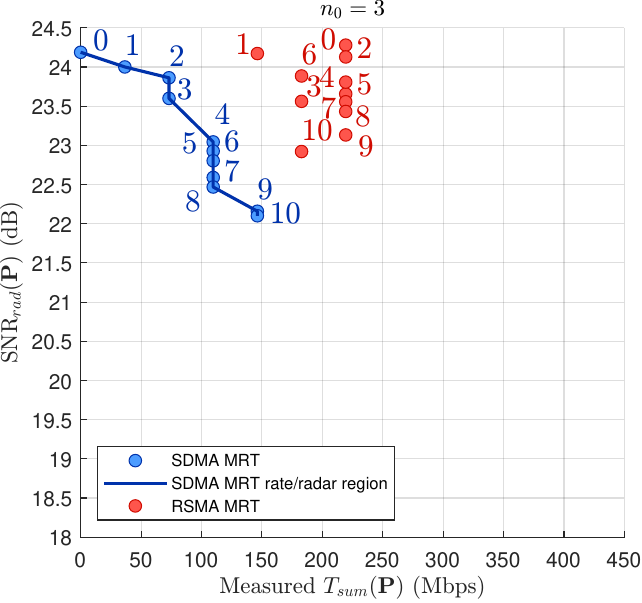}
  \caption{S3 (MRT)}
\end{subfigure}
\\
\begin{subfigure}{0.3\textwidth}
  \includegraphics[width=\linewidth]{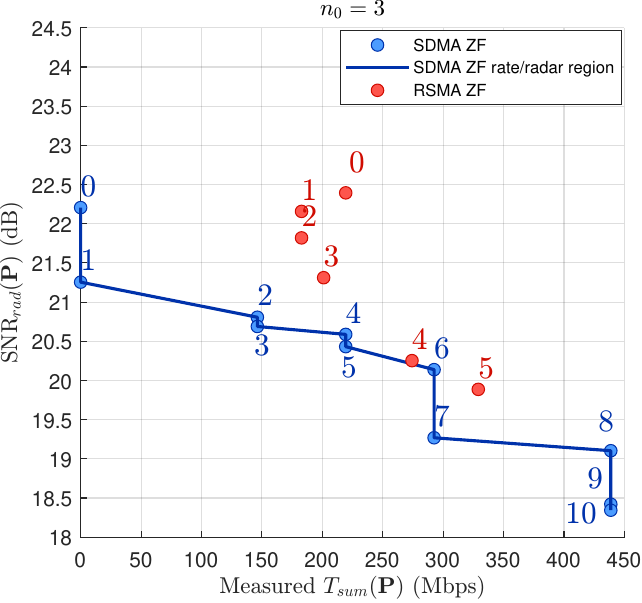}
  \caption{S1 (ZF)}
\end{subfigure}\hfil % <-- added
\begin{subfigure}{0.3\textwidth}
  \includegraphics[width=\linewidth]{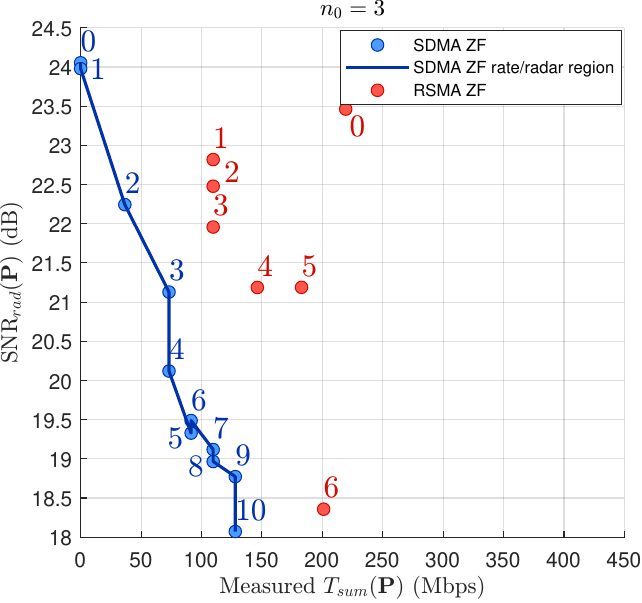}
  \caption{S2 (ZF)}
\end{subfigure}\hfil % <-- added
\begin{subfigure}{0.3\textwidth}
  \includegraphics[width=\linewidth]{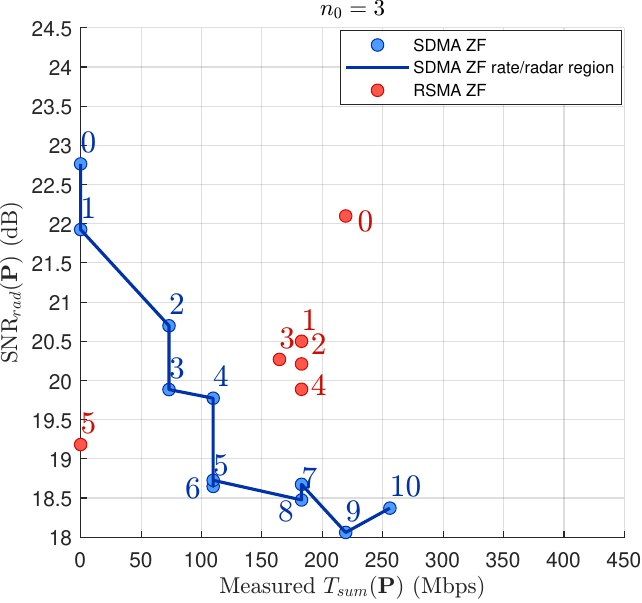}
  \caption{S3 (ZF)}
  \end{subfigure}
\caption{The measured ISAC performance region for RSMA and SDMA, for the precoder parameters in Table~\ref{tab:parameter_both}. The MCS levels in Table~\ref{tab:parameter_both} are used as a starting point and progressively decreased/increased the MCS levels in case of any/no decoding errors. Hence, any difference between the expected $T_{\rm sum}(\nbP)$ in Fig.~\ref{fig:sim} and the measured $T_{\rm sum}(\nbP)$ in Fig.~\ref{fig:isac_region_measure} boils down to a mismatch between the initial MCS levels in Table~\ref{tab:parameter_both} and ones that maximized the measured sum throughput.}
\label{fig:isac_region_measure}
\end{figure*}

For the target situated at $5.25{\rm m}~(n_0 = 3)$, Fig.~\ref{fig:isac_region_measure} plots the measured ISAC performance region for the precoder parameters parameters listed in Table~\ref{tab:parameter_both}. To measure $T_{\rm sum}(\nbP)$, we identified the highest MCS levels that resulted in error-free decoding of the five OFDM symbols in the data payload (see Fig.~\ref{fig:ofdm signal stages}) by using the MCS levels in Table~\ref{tab:parameter_both} as a starting point and progressively decreasing/increasing the MCS levels if there were any/no decoding errors. Hence, any difference between the expected $T_{\rm sum}(\nbP)$ in Fig.~\ref{fig:sim} and the measured $T_{\rm sum}(\nbP)$ in Fig.~\ref{fig:isac_region_measure} boils down to a mismatch between the initial MCS levels in Table~\ref{tab:parameter_both}-b and ones that maximized the measured sum throughput.

The following aspects of Fig.~\ref{fig:sim} are validated by Fig.~\ref{fig:isac_region_measure}:
\begin{itemize}
    \item \emph{RSMA ISAC \textbf{empirically} outperforms SDMA ISAC}: Except for point 5 in S1 (MRT) and point 5 in S3 (ZF), all other RSMA measurement points lie to the northeast of the SDMA ISAC performance boundary. Consistent with the heatmaps from Fig.~\ref{fig:ranging_heatmaps}, the measured ${\sf SNR}_{\rm rad}(\nbP)$ is comparable for both MRT-based SDMA and MRT-based RSMA, varying between $20$ to $24{\rm dB}$. 
    The gain in peak throughput for RSMA over SDMA is:
   \begin{itemize}
            \item S1 MRT-based: $-6\%$ (SDMA: $292.5{\rm Mbps}$, RSMA: $274{\rm Mbps}$)
            \item S1 ZF-based: $-33.3\%$ (SDMA: $438.75{\rm Mbps}$, RSMA: $329.06{\rm Mbps}$)
            \item S2 MRT-based: $12\%$ (SDMA: $146.25{\rm Mbps}$, RSMA: $164{\rm Mbps}$)
            \item S2 ZF-based: $41.5\%$ (SDMA: $128{\rm Mbps}$, RSMA: $219.4{\rm Mbps}$)
            \item S3 MRT-based: $50\%$ (SDMA: $146.25{\rm Mbps}$, RSMA: $219{\rm Mbps}$)
            \item S3 ZF-based: $-16\%$ (SDMA: $256{\rm Mbps}$, RSMA: $219.4{\rm Mbps}$)
    \end{itemize} 
    Despite the caveat that our measured throughput is based on the decoding of only five OFDM symbols, the MRT-based SDMA experiences a drop of more than $50\%$ in sum throughput in Scenarios S2 and S3 compared to Scenario S1. This is due to high inter-user interference, and the results are consistent with Fig.~\ref{fig:sim}. ZF-based SDMA shows larger sum throughput compared to MRT-based SDMA in S1 and S3. However, ZF-based SDMA shows no improvement in sum throughput in S2, indicating that the inter-user interference in S2 is not manageable for both ZF and MRT precoder designs.
    
    \item \emph{RSMA throughput collapse due to imperfect SIC and finite MCS levels}: For point 5 (RSMA) in S1 (MRT) and point 5 (RSMA) in S3 (ZF), the common stream cannot be decoded even at the lowest MCS level (BPSK rate 1/2), possibly due to large CSI estimation errors. Hence, imperfect SIC causes throughput collapse.
\end{itemize}
% The SDMA ISAC region seen in the Fig.~\ref{fig:sim} is mirrored in the measurements as shown in Fig.~\ref{fig:isac_region_measure}. The discretized achievable throughput (measured $T_{sum}$) aligns some of the results in the same line, even though they have different parameter settings and radar SNR performance.

Finally, the main aspect in which Fig.~\ref{fig:isac_region_measure} differs from Fig.~\ref{fig:sim} is the absence of a clear RSMA performance boundary. This is due to the mismatch between the MCS levels in Table~\ref{tab:parameter_both}-c and the ones that maximize the measured sum throughput for a given RSMA point. In particular, if the common stream cannot be decoded at the MCS level in Table~\ref{tab:parameter_both}-c, it results in a lower measured sum throughput than in Fig.~\ref{fig:sim}. An extreme example of this is RSMA points 5 in both Fig.~\ref{fig:isac_region_measure}a and Fig.~\ref{fig:isac_region_measure}f, where the common stream cannot be decoded at the lowest MCS level ($\ncalM_c = 0$), which in turn causes throughput collapse. Since the measured RSMA throughput is based on the decoding of only five OFDM symbols per stream, there is no systematic pattern to this MCS level mismatch. This issue can be addressed through larger payloads and link adaptation, yielding a long-term average measured throughput that should make the RSMA performance boundary clearer. This can be the subject of ongoing work.

%\begin{itemize}
%      \item      
      %To understand the reason for the throughput degradation, we need to look at the parameters Table~\ref{tab:parameter_both} in measuring points 0 and 1. The main difference between those two measurements is in measuring point 0, there is no SIC involved since there is a common stream only ($t_p=0$), the transmitter is broadcasting the common stream only to both UEs, and both UEs are only required to decode the common stream, compared to measurements in point 1, private streams are activated ($t_p\neq 0$), both UEs are required to perform SIC to decode both common stream and the corresponding private stream. The reconstruction of the common stream symbols of UE $i$ is based on the estimation of $\nbh_i^H\nbp_c$, the imperfect estimated precoded CSI causes imperfection\footnote{This imperfection in SIC at UE $i$ is caused by estimation error of precoded CSI ($\hat{\nbh}_i^H\nbp_c \approx\nbh_i^H\nbp_c$) which the main imperfection is introduced in the reconstruction procedure and channel estimation. However, the imperfect SIC mentioned in Section~\ref{subsec:sim_param_search} is caused by the unsuccessful decoded common stream ($\hat{s}_c\neq \hat{s}$), which is introduced in the decoding common stream procedure.}. The common stream can propagate residual interference to subsequent streams (private streams). This cascading effect may degrade the decoding performance due to the increase of SINR.

\section{Conclusion}
\label{sec: concl}
In this paper, we presented the first ever experimental study of RSMA ISAC using SDRs. We started by formulating signal model that included four different ISAC precoder design choices -- RSMA/SDMA with/without dedicated sensing signal -- as special cases. Using sum throughput as the communications metric and (post-processing) radar SNR as the sensing metric, we defined the ISAC performance region to evaluate these design choices. Over three measurement scenarios that are representative of vehicular ISAC, involving communicating with two users and ranging a single target with different levels of inter-user interference and separation/integration between sensing and communications, we observed that RSMA has a larger ISAC performance region than SDMA. Notably, RSMA without a dedicated sensing signal outperformed SDMA with a dedicated sensing signal, by achieving peak sum throughput gains of  up to $50\%$ for similar radar SNR (between $20$ and $24{\rm dB}$).

Though further enhancement of this prototype is possible by incorporating link adaptation, our results provide compelling answers to questions Q1 and Q2 posed on page 1 -- i.e., \emph{RSMA without a dedicated sensing signal achieves better ISAC performance than SDMA with a dedicated sensing signal}.

\begin{comment}
  \section*{Acknowledgment}
The authors would like to thank Guoqian Sun, Steven Zhang and Ziang Liu for their assistance during the measurement.  
\end{comment}

\appendices
\section{Phase Offset Correction}
\label{appendix:A}
To account for any mismatch between the desired and the true steering vector of the TX array, we perform calibration as per \cite{Radar_conf_2020_doa} by placing an \emph{anchor} antenna at the broadside of the TX array. Each array element is sequentially activated to transmit a reference signal sequence that is captured by the anchor. The measured channel between the anchor and the $g$-th TX antenna element ($g \in \{0,1\}$) over subcarrier $k$ is given by:
\begin{align}
    h_{g}[k] = \beta \underbrace{\mathrm{exp}\left(j2\pi\frac{n_a}{N_c}k\right)}_\text{Propagation delay}\cdot \underbrace{\mathrm{exp}\left(j\phi_{g,k}\right)}_\text{RF chain}\cdot\underbrace{\mathrm{exp}\left(j2\pi\frac{dg}{\lambda} \sin \theta_a\right)}_\text{Extra propagation delay},
\end{align}
where $\beta$ denotes the signal attenuation and the first phase term is due to the propagation delay. The second phase term is introduced by the RF chain hardware \cite{Trans_instrument_radar_goverdovsky}, and the last term is the additional phase shift for a TX element relative to the reference element ($g = 0$), which depends on the direction of arrival $\theta_a$ ($0$ for the broadside). 

Let $\phi_{g}[k]$ denote the phase of $h_g[k]$. Then, $\Delta \phi=\frac{1}{N_c}\sum\limits_{k=0}^{N_c-1}(\phi_{0}[k]-\phi_{1}[k])$ captures the phase difference (averaged over all subcarriers) of the channels between the TX array elements and the anchor. Ideally, $\Delta \phi$ should be equal to zero for an anchor at the array broadside. Hence, as calibration, we apply a phase shift of $-\Delta \phi$ to the signal transmitted out of the second TX element ($g = 1$).

\section{The Cramér-Rao Bound of Ranging Delay Estimation}
\label{appendix:B}
First, we simplified (\ref{eq: reflect signal}) into $y[k]=s(k;n_0)+n_r[k]$, where $s(k;n_0)=\beta\mathbf{a}_0^H\mathbf{x}[k]\cdot \mathrm{exp}(j2\pi\frac{n_0}{N_c}k)$. Then the log-likelihood is:
\begin{align}
    \label{eq:log likelihood}
    \log~p(y;n_0)=-\frac{1}{\sigma^2}\sum_{k=0}^{N_c-1}|y[k]-s(k;n_o)|^2+\mathrm{const}.
\end{align}
By differentiating the log-likelihood function, we can get
\begin{align}
\label{eq:log likelihood derive 1}
    \frac{\partial \log~p}{\partial n_0}=\frac{2}{\sigma^2}\Re \Bigl\{\sum_k(y[k]-s(k;n_0))^*\frac{\partial s(k;n_0)}{\partial n_0}\Bigr\},
\end{align}
in which $\frac{\partial s(k;n_0)}{\partial n_0}=\frac{j2\pi k}{N_c}s(k;n_0)$. 
The Fisher Information can be written as:
\begin{align}
    \label{eq:fisher information}
    \mathcal{I}(n_0)&=\mathbb{E}\Bigl[\Bigl(\frac{\partial \log~p}{\partial n_0} \Bigr)^2 \Bigr] \\
    &=\frac{2}{\sigma^2}\sum^{N_c-1}_{k=0}\Bigl|\frac{\partial s(k;n_0)}{\partial n_0}\Bigr|^2\\
   % &=\frac{8\pi^2}{\sigma^2 N_c^2}\sum_{k=0}^{N_c-1}k^2s(k;n_0)^2\\
    &=\frac{8\pi^2\beta^2}{\sigma^2N_c^2}\sum_{k=0}^{N_c-1}k^2|\mathbf{a}_0^H\mathbf{x}[k]|^2.
\end{align}
Then (\ref{eq:CRB}) can be calculated by inverting the Fisher Information in (\ref{eq:fisher information}), which is
\begin{align}
    \label{eq:invert fisher info}
    {\sf CRB}(n_0)&= \frac{1}{\mathcal{I}(n_0)}=\frac{\sigma^2N_c^2}{8\pi^2\beta^2\sum_{k=0}^{N_c-1}k^2|\mathbf{a}_0^H\mathbf{x}[k]|^2}.
\end{align}

%Each element in the TX array experiences unique phase noise due to manufacturing variability. To ensure an accurate steering vector for the array, it is necessary to calibrate these elements so that there is no phase difference before the measurement. This calibration process in \cite{Radar_conf_2020_doa} requires placing an anchor antenna at the broadside of the transmitter array. Each array element is sequentially activated to transmit a reference signal sequence to the anchor receiver, allowing the measurement of the channel between each transmitter element and the anchor antenna individually. The measured channel between transmitter antenna-$g$ to anchor on $k$-th subcarrier can be denoted as:

%The calibration process will estimate the random phase difference caused by the RF components between TX-$0$ and TX-$1$ (), then the estimation of phase difference $\Delta\hat{\phi}$ is applied on TX-$1$ to compensate the phase difference.

\bibliographystyle{IEEEtran}

\end{document}

%% file: notation_new.tex
\def\nba{{\mathbf{a}}}

\def\nbh{{\mathbf{h}}}

\def\nbp{{\mathbf{p}}}

\def\nbu{{\mathbf{u}}}

\def\nbx{{\mathbf{x}}}

\def\nb0{{\mathbf{0}}}
\def\nb1{{\mathbf{1}}}

\def\nbH{{\mathbf{H}}}

\def\nbP{{\mathbf{P}}}

\def\nbU{{\mathbf{U}}}

\def\ncalC{{\mathcal{C}}}

\def\ncalM{{\mathcal{M}}}
\def\ncalN{{\mathcal{N}}}

\def\nbbC{{\mathbb{C}}}

\def\nbbM{{\mathbb{M}}}

\newtheorem{ndef}{Definition}
\newtheorem{nrem}{Remark}

\newtheorem{eg}{Example}

%% file: paper_revised_r2.bbl
\begin{thebibliography}{10}
\providecommand{\url}[1]{#1}
\csname url@samestyle\endcsname
\providecommand{\newblock}{\relax}
\providecommand{\bibinfo}[2]{#2}
\providecommand{\BIBentrySTDinterwordspacing}{\spaceskip=0pt\relax}
\providecommand{\BIBentryALTinterwordstretchfactor}{4}
\providecommand{\BIBentryALTinterwordspacing}{\spaceskip=\fontdimen2\font plus
\BIBentryALTinterwordstretchfactor\fontdimen3\font minus \fontdimen4\font\relax}
\providecommand{\BIBforeignlanguage}[2]{{%
\expandafter\ifx\csname l@#1\endcsname\relax
\typeout{** WARNING: IEEEtran.bst: No hyphenation pattern has been}%
\typeout{** loaded for the language `#1'. Using the pattern for}%
\typeout{** the default language instead.}%
\else
\language=\csname l@#1\endcsname
\fi
#2}}
\providecommand{\BIBdecl}{\relax}
\BIBdecl

\bibitem{3gpp_isac}
``Technical {S}pecification {G}roup {TSG SA}; {S}ervice requirement for {I}ntegrated {S}ensing and {C}ommunications (release 19),'' {TS} 22.137 (v0.1.0), 3{GPP}, Aug. 2023.

\bibitem{Liu_jsac_survey}
F.~Liu \emph{et~al.}, ``Integrated sensing and communications: Toward dual-functional wireless networks for {6G} and beyond,'' \emph{{IEEE} J. Sel. Areas Commun.}, vol.~40, no.~6, pp. 1728--1767, 2022.

\bibitem{PIEEE_bruno_2024_ma4intelligent6g}
B.~Clerckx \emph{et~al.}, ``{M}ultiple {A}ccess {T}echniques for {I}ntelligent and {M}ultifunctional {6G}: {T}utorial, {S}urvey, and {O}utlook,'' \emph{Proceedings of the IEEE}, vol. 112, no.~7, pp. 832--879, 2024.

\bibitem{liuXiang_TSP_MIMOprecoder}
X.~Liu \emph{et~al.}, ``Joint transmit beamforming for multiuser {MIMO} communications and {MIMO} radar,'' \emph{{IEEE} Trans. Signal Process.}, vol.~68, pp. 3929--3944, 2020.

\bibitem{Liu_TSP_MinCRB_QoS}
F.~Liu \emph{et~al.}, ``Cramér-{R}ao bound optimization for joint radar-communication beamforming,'' \emph{{IEEE} Trans. Signal Process.}, vol.~70, pp. 240--253, 2022.

\bibitem{Lu_TSP_dedicated_required}
S.~Lu \emph{et~al.}, ``Random {ISAC} signals deserve dedicated precoding,'' \emph{{IEEE} Trans. Signal Process.}, vol.~72, pp. 3453--3469, 2024.

\bibitem{cui_wcnc_MSE_DoF}
Y.~Cui \emph{et~al.}, ``Optimal precoding design for monostatic {ISAC} systems: {MSE} lower bound and {DoF} completion,'' in \emph{Proc. of the IEEE Wireless Commun. and Networking Conf. (WCNC)}, 2024, pp. 1--6.

\bibitem{Liu_erzi_ISAC_SI_JSAC_2023}
Z.~Liu, S.~Aditya, H.~Li, and B.~Clerckx, ``Joint transmit and receive beamforming design in full-duplex integrated sensing and communications,'' \emph{{IEEE} J. Sel. Areas Commun.}, vol.~41, no.~9, pp. 2907--2919, 2023.

\bibitem{Fan_mMIMO_isac_iot_2024}
Y.~Fan, S.~Wu, X.~Bi, and G.~Li, ``Power allocation for cell-free {M}assive {MIMO} {ISAC} systems with {OTFS} signal,'' \emph{{IEEE} Internet Things J.}, pp. 1--1, 2024.

\bibitem{liao_mMIMO_isac_twc_2024}
B.~Liao, H.~Q. Ngo, M.~Matthaiou, and P.~J. Smith, ``Power allocation for massive {MIMO}-{ISAC} systems,'' \emph{{IEEE} Trans. Wireless Commun.}, vol.~23, no.~10, pp. 14\,232--14\,248, 2024.

\bibitem{topal_mMIMO_isac_wcml_2024}
O.~A. Topal,  .~T. Demir, E.~Björnson, and C.~Cavdar, ``Multi-target integrated sensing and communications in massive {MIMO} systems,'' \emph{{IEEE} Wireless Commun. Lett.}, pp. 1--1, 2024.

\bibitem{demirhan_ML_isac_mMIMO_spawc_2024}
U.~Demirhan and A.~Alkhateeb, ``Learning beamforming in cell-free {M}assive {MIMO} {ISAC} systems,'' in \emph{2024 IEEE 25th International Workshop on Signal Processing Advances in Wireless Communications (SPAWC)}, 2024, pp. 326--330.

\bibitem{Liu_TWC_radar_predict_vehiculat}
F.~Liu, W.~Yuan, C.~Masouros, and J.~Yuan, ``Radar-assisted predictive beamforming for vehicular links: Communication served by sensing,'' \emph{{IEEE} Trans. Wireless Commun.}, vol.~19, no.~11, pp. 7704--7719, 2020.

\bibitem{Yuan_TWC_V2I_vehicular}
W.~Yuan \emph{et~al.}, ``Bayesian predictive beamforming for vehicular networks: {A} low-overhead joint radar-communication approach,'' \emph{{IEEE} Trans. Wireless Commun.}, vol.~20, no.~3, pp. 1442--1456, 2021.

\bibitem{qi_TCOML_ISAC_HBF_mmWave}
C.~Qi, W.~Ci, J.~Zhang, and X.~You, ``Hybrid beamforming for millimeter wave {MIMO} integrated sensing and communications,'' \emph{{IEEE} Commun. Lett.}, vol.~26, no.~5, pp. 1136--1140, 2022.

\bibitem{wang_TCOM_2022_ISAC_HybridBF}
X.~Wang, Z.~Fei, J.~A. Zhang, and J.~Xu, ``Partially-connected hybrid beamforming design for integrated sensing and communication systems,'' \emph{{IEEE} Trans. Commun.}, vol.~70, no.~10, pp. 6648--6660, 2022.

\bibitem{gupta_mmwave_ISAC_OJ_2024}
A.~Gupta \emph{et~al.}, ``An affine precoded superimposed pilot-based mmwave {MIMO}-{OFDM} {ISAC} system,'' \emph{IEEE Open Journal of the Communications Society}, vol.~5, pp. 1504--1524, 2024.

\bibitem{wang_mmwave_isac_iot_2024}
P.~Wang \emph{et~al.}, ``Low-complexity joint transceiver optimization for {MmWave/THz} {MU}-{MIMO} {ISAC} systems,'' \emph{{IEEE} Internet Things J.}, pp. 1--1, 2024.

\bibitem{Lin_mmwave_isac_onebit_2024}
Q.~Lin \emph{et~al.}, ``One-bit transceiver optimization for mmwave integrated sensing and communication systems,'' \emph{{IEEE} Trans. Commun.}, pp. 1--1, 2024.

\bibitem{lyu2023prototype}
X.~Lyu, S.~Aditya, J.~Kim, and B.~Clerckx, ``{R}ate-{S}plitting {M}ultiple {A}ccess: The {F}irst {P}rototype and {E}xperimental {V}alidation of its {S}uperiority over {SDMA} and {NOMA},'' \emph{{IEEE} Trans. Wireless Commun.}, vol.~23, no.~8, Aug. 2024.

\bibitem{RSMA_MGM_prototype}
X.~Lyu, S.~Aditya, and B.~Clerckx, ``Rate-{S}plitting {M}ultiple {A}ccess for overloaded multi-group multicast: A first experimental study,'' \emph{{IEEE} Trans. Broadcast.}, Oct. 2024, (Early Access).

\bibitem{RSMA_JSAC_Primer}
B.~Clerckx \emph{et~al.}, ``A {P}rimer on {R}ate-{S}plitting {M}ultiple {A}ccess: {T}utorial, {M}yths, and {F}requently {A}sked {Q}uestions,'' \emph{{IEEE} J. Sel. Areas Commun.}, vol.~41, no.~5, pp. 1265--1308, May 2023.

\bibitem{Adi_ccs_ojcoms_2023}
S.~Aditya, O.~Dizdar, B.~Clerckx, and X.~Li, ``Sensing using coded communications signals,'' \emph{IEEE Open Journal of the Communications Society}, vol.~4, pp. 134--152, 2023.

\bibitem{XuCC_JTSP_RSMA_ISAC}
C.~Xu \emph{et~al.}, ``Rate-splitting multiple access for multi-antenna joint radar and communications,'' \emph{{IEEE} J. Sel. Topics Signal Process.}, vol.~15, no.~6, pp. 1332--1347, 2021.

\bibitem{Liu_ICC_RSMA_ISAC_2023}
Z.~Liu, Y.~Jint, B.~Cao, and R.~Lu, ``{RISAC}: {R}ate-{S}plitting {M}ultiple {A}ccess enabled {I}ntegrated {S}ensing and {C}ommunication systems,'' in \emph{Proc. of the IEEE Intl. Conf. on Commun. (ICC)}, 2023, pp. 6449--6454.

\bibitem{Liu_TWC_RSMA_ISAC_sate}
Z.~Liu, L.~Yin, W.~Shin, and B.~Clerckx, ``Rate-splitting multiple access for quantized {ISAC} {LEO} satellite systems: A max-min fair energy-efficient beam design,'' \emph{{IEEE} Trans. Wireless Commun.}, vol.~23, no.~10, pp. 15\,394--15\,408, 2024.

\bibitem{loli2022ratesplitting}
\BIBentryALTinterwordspacing
R.~C. Loli, O.~Dizdar, and B.~Clerckx, ``{R}ate-{S}plitting {M}ultiple {A}ccess for multi-antenna joint radar and communications with partial {CSIT}: {P}recoder {O}ptimization and {L}ink-{L}evel simulations,'' 2022. [Online]. Available: \url{https://arxiv.org/abs/2201.10621}
\BIBentrySTDinterwordspacing

\bibitem{YinLF_CommL_RSMA_ISAC_wo}
L.~Yin, Y.~Mao, O.~Dizdar, and B.~Clerckx, ``Rate-splitting multiple access for {6G}—{Part II}: Interplay with integrated sensing and communications,'' \emph{{IEEE} Commun. Lett.}, vol.~26, no.~10, pp. 2237--2241, 2022.

\bibitem{ISAC_SISO_proto_1}
S.~Ding \emph{et~al.}, ``{I}ntegrated {S}ensing and {C}ommunication: {P}rototype and {K}ey {P}rocessing {A}lgorithms,'' in \emph{Proc. of the IEEE Intl. Conf. on Commun. (ICC) Workshops}, 2023, pp. 225--230.

\bibitem{ISAC_SISO_proto_2}
------, ``Channel measurements for integrated sensing and communication: Method and prototype test,'' in \emph{Proc. of the 99th IEEE Vehicular Tech. Conf. (VTC-Spring)}, 2024, pp. 01--06.

\bibitem{ISAC_SISO_proto_otfs}
M.~Ouyang, Y.~Wang, T.~Zhang, and F.~Gao, ``An {RF}{S}o{C}-based scalable {ISAC} prototyping platform with {OTFS} waveform,'' in \emph{Proc. of the IEEE Wireless Commun. and Netw. Conf. (WCNC)}, 2024, pp. 1--5.

\bibitem{Xu_OJ_2022_ISAC_MIMO_proto_beamform}
T.~Xu, F.~Liu, C.~Masouros, and I.~Darwazeh, ``An experimental proof of concept for integrated sensing and communications waveform design,'' \emph{IEEE Open Journal of the Communications Society}, vol.~3, pp. 1643--1655, 2022.

\bibitem{IEEE_80211_ac}
``{IEEE} {S}tandard for {I}nformation technology - {T}elecommunications and information exchange between systems - local and metropolitan area networks -specific requirements - part 11: {W}ireless {LAN} {M}edium {A}ccess {C}ontrol ({MAC}) and {P}hysical {L}ayer ({PHY}) {S}pecifications--{A}mendment 4: Enhancements for very high throughput for operation in bands below 6 {GH}z.'' {IEEE} {S}td 802.11ac ({TM}) - 2013, pp. 1--425, 2013.

\bibitem{RSMA_NOUM_prototype}
X.~Lyu, S.~Aditya, and B.~Clerckx, ``Rate-splitting {M}ultiple {A}ccess for {N}on-{O}rthogonal {U}nicast {M}ulticast: {A}n experimental study,'' in \emph{Proc. of the 25th IEEE Intl. Workshop on Signal Proc. Advances in Wireless Commun. (SPAWC)}, 2024, pp. 591--595.

\bibitem{Shannon_bound_adjustment}
P.~Mogensen \emph{et~al.}, ``{LTE} capacity compared to the {S}hannon bound,'' in \emph{Proc. of the 65th IEEE Vehicular Tech. Conf. (VTC Spring)}, 2007, pp. 1234--1238.

\bibitem{Rafael_spawc_2021}
R.~Cerna-Loli, O.~Dizdar, and B.~Clerckx, ``A rate-splitting strategy to enable joint radar sensing and communication with partial csit,'' in \emph{Proc. of the 22nd IEEE Intl. Workshop on Signal Proc. Advances in Wireless Commun. (SPAWC)}, 2021, pp. 491--495.

\bibitem{HamdiMISOImperfectCSIT}
H.~Joudeh and B.~Clerckx, ``Sum-rate maximization for linearly precoded downlink multiuser {MISO} systems with partial {CSIT}: {A} rate-splitting approach,'' \emph{{IEEE} Trans. Commun.}, vol.~64, no.~11, pp. 4847--4861, 2016.

\bibitem{balanis2016antenna}
C.~A. Balanis, \emph{Antenna theory: analysis and design}.\hskip 1em plus 0.5em minus 0.4em\relax John wiley \& sons, 2016.

\bibitem{FinephaseShifting}
H.~Minn, ``A robust timing and frequency synchronization for {OFDM} systems,'' \emph{{IEEE} Trans. Wireless Commun.}, vol.~2, no.~4, pp. 822--839, 2003.

\bibitem{trifonovPolar}
P.~Trifonov, ``{E}fficient {D}esign and {D}ecoding of {P}olar {C}odes,'' \emph{{IEEE} Trans. Commun.}, vol.~60, no.~11, pp. 3221--3227, 2012.

\bibitem{constructionPolar}
H.~Li and J.~Yuan, ``{A} {P}ractical {C}onstruction {M}ethod for {P}olar {C}odes in {AWGN} {C}hannels,'' in \emph{IEEE 2013 Tencon - Spring}, 2013, pp. 223--226.

\bibitem{listdecoding}
I.~Tal and A.~Vardy, ``{L}ist {D}ecoding of {P}olar {C}odes,'' \emph{{IEEE} Trans. Inf. Theory}, vol.~61, no.~5, pp. 2213--2226, 2015.

\bibitem{Radar_conf_2020_doa}
L.~Storrer \emph{et~al.}, ``Experimental implementation of a multi-antenna 802.11ax-based passive bistatic radar for human target detection,'' in \emph{Proc. of the IEEE Radar Conf. (RadarConf20)}, 2020, pp. 1--6.

\bibitem{Trans_instrument_radar_goverdovsky}
V.~Goverdovsky \emph{et~al.}, ``Modular {S}oftware-{D}efined {R}adio testbed for rapid prototyping of localization algorithms,'' \emph{{IEEE} Trans. Instrum. Meas.}, vol.~65, no.~7, pp. 1577--1584, 2016.

\end{thebibliography}
